\documentclass{article}

\usepackage{arxiv}
\usepackage{natbib}
\usepackage[utf8]{inputenc} 
\usepackage[T1]{fontenc}    
\usepackage{url}            
\usepackage{booktabs}       
\usepackage{amsfonts}       
\usepackage{nicefrac}       
\usepackage{microtype}      
\usepackage{lipsum}         
\usepackage{graphicx}
\usepackage{doi}
\usepackage[inline]{enumitem}
\usepackage{amsmath}
\usepackage{amsthm}         

\DeclareMathOperator*{\argmax}{arg\,max}

\usepackage{subcaption}     
\usepackage{algorithm}
\usepackage{algorithmic}

\usepackage{hyperref}       
\usepackage{cleveref}       

\title{Robust Self-Triggered Control Approaches Optimizing Sampling Sequences with Synchronous Measurements}

\newif\ifuniqueAffiliation
\uniqueAffiliationtrue

\ifuniqueAffiliation 
\author{ \href{https://orcid.org/0000-0001-5024-9224}{\includegraphics[scale=0.06]{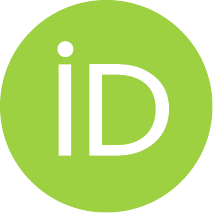}\hspace{1mm}Abbas Tariverdi}\thanks{This work was conducted while Abbas Tariverdi was with CRIStAL (Research Center in Computer Science, Signal and Automatic Control of Lille), UMR 9189, CNRS, University of Lille, Centrale Lille, Lille, France.} \\
	\texttt{abbasta@abbasta.com} 
}
\else
\usepackage{authblk}

\newbox{\orcid}\sbox{\orcid}{\includegraphics[scale=0.06]{orcid.pdf}} 
\author[1]{%
	\href{https://orcid.org/0000-0000-0000-0000}{\usebox{\orcid}\hspace{1mm}David S.~Hippocampus\thanks{\texttt{hippo@cs.cranberry-lemon.edu}}}%
}
\author[1,2]{%
	\href{https://orcid.org/0000-0000-0000-0000}{\usebox{\orcid}\hspace{1mm}Elias D.~Striatum\thanks{\texttt{stariate@ee.mount-sheikh.edu}}}%
}
\affil[1]{Department of Computer Science, Cranberry-Lemon University, Pittsburgh, PA 15213}
\affil[2]{Department of Electrical Engineering, Mount-Sheikh University, Santa Narimana, Levand}
\fi

\renewcommand{\shorttitle}{Self-Triggered Controls Optimizing Sampling Sequences with Synchronous Measurements}

\newtheorem{proposition}{Proposition}

\newtheorem{remark}{Remark} 
\newtheorem{assumption}{Assumption}    
\newtheorem{definition}{Definition}    

\hypersetup{
pdftitle={A template for the arxiv style},
pdfsubject={q-bio.NC, q-bio.QM},
pdfauthor={David S.~Hippocampus, Elias D.~Striatum},
pdfkeywords={First keyword, Second keyword, More},
}

\begin{document}
\maketitle

\begin{abstract}
\textit{Note: This research was conducted in 2017--2018. The literature review has not been updated and may not reflect subsequent or concurrent developments in the field.} 

Feedback control algorithms traditionally rely on periodic execution on digital platforms. While this simplifies design and analysis, it often leads to inefficient resource usage (e.g., CPU, network bandwidth) in embedded control and shared networks. This work investigates self-triggering implementations of linear controllers in sampled-data systems with synchronous measurements. Our approach precomputes the next sampling sequence over a finite horizon based on current state information. We introduce a novel optimal self-triggering scheme that guarantees exponential stability for unperturbed systems and global uniform ultimate boundedness for perturbed systems. This ensures robustness against external disturbances with explicit performance guarantees. Simulations demonstrate the benefits of our approach.
\end{abstract}


\section{Introduction}
Due to the fast development of computers, the majority of control systems are nowadays implemented on digital hardware. Growing popularity of Networked Control Systems (NCSs) demands economic use of computational and communication resources and explicitly addressing of energy, computation, and communication constraints when designing feedback control loops \citep{heemels2012introduction}. The response to these challenges implies re-answering to the following questions "when to sample?", "when
to update control action?” or "when to transmit information?" \citep{miskowicz2018event}.

Event-based scheme is a general solution to these answers. Although event-based control (including event-triggered and self-triggered control) offers some clear advantages with respect to periodic control, it introduces some new theoretical and practical problems. Main element of event-triggered and self-triggered control systems a triggering mechanism that determines when the control input has to be updated again. In event-triggered control a triggering condition based on current measurements is continuously monitored and when violated, an event is triggered. In self-triggered control the next update time is precomputed using last sampled measurements and knowledge on the plant dynamics. 

To sum up in a simple way, the following problem has been of a great importance: the stability of sampled-data systems with a self-triggering mechanism aims at enlarging the average of sampling intervals while reducing the computational and energetic costs. The work presented here is concerned with the aforementioned problem. The main objective is to design a self-triggering mechanism allows for enlarging the average of sampling intervals of state-feedback control for linear sampled-data systems while ensuring the system stability.

In the recent years, considerable amount of research has gone into the stability analysis of sampled-data systems. The authors in \citep{hetel2017recent} have reviewed concepts and approaches in the stability of sampled-data systems with deterministic aperiodic sampling. They proposed a structural framework for stability analysis of the systems. The followings are main perspectives for addressing the stability analysis:
\begin{enumerate*}
	\item Time delay approach,
	\item Hybrid system approach,
	\item Discrete-time approach and convex-embeddings,
	\item Input/output stability approach.
\end{enumerate*}

The time delay approach in stability analysis was initially introduced in \citep{mikheev1988asymptotic,aastrom2013adaptive}, and later developed in \citep{fridman1992use,louisell2001delay,teel1998note}. The underlying approach is that the LTI system is remodeled as an LTI system with a time varying delay, which enables to accommodate fast varying delays \citep{fridman2003delay,niculescu2012advances}.

The hybrid systems approach to stability analysis of LTI systems with uniform and multi
rate sampling, was developed in \citep{sun1993h,kabamba1993worst}. Stability analysis under aperiodic sampling,
using the hybrid systems approach was studied using impulsive models \citep{toivonen1992sampled,dullerud1999asynchronous}. 

In the discrete-time systems approach, stability criteria have been put forward by analyzing the spectral radius \citep{blondel2005computationally}, or by validating the existence of Lyapunov functions that are quasi-quadratic \citep{molchanov1989criteria}, parameter dependent \citep{daafouz2001parameter}, path dependent \citep{lee2006uniform}, non-monotonic \citep{megretski1996integral}, and composite quadratic \citep{hu2010non}. The discrete-time approach has also been employed in tracking control of sampled-data systems with uncertain time varying delay \citep{van2010tracking}. Further study using this approach has also gone into incorporating phenomena like packet dropouts, time-varying sampling intervals and time varying delays larger than the sampling interval \citep{cloosterman2010controller}.

The final approach, namely the input-output approach considers sampling and delay as perturbations with respect to the nominal continuous-time control loop. Powerful stability criteria using the input-output stability approach were proposed for LTI systems with time-varying bounded delays in \citep{kao2004simple}. Another important result in this direction was provided in \citep{mirkin2007some}, where LTI systems with aperiodic sampling was considered and results obtained in \citep{fridman2004robust} (using Lyapunov-Krasovskii approach) are derived alternately using scaled small gain conditions. Further improvement of these methods using dissipativity, anti-passivity, and IQCs have been proposed in \citep{fujioka2009stability,omran2014stability}.

In self-triggered control the next update time is precomputed at a control update time based on predictions using previously received data and knowledge on the plant dynamics where there is no requirement to continuously evaluate the triggering condition. A first attempt to explore self-triggered models for linear systems was developed
	in \citep{velasco2003self}, by discretizing the plant, and in \citep{wang2008state} for linear $ H_{\infty} $ controllers. The authors in \citep{velasco2003self} addressed a self triggering mechanism based on CPU utilization factors and control performances. This paper and references therein can be useful for real-time scheduling of event-/self-triggered control tasks.  
	
	Self-triggered control was developed for state-dependent homogeneous and polynomial systems in \citep{anta2008self} and \citep{anta2010sample}. The event-triggering scheme is as the mechanism that was developed in \citep{tabuada2007event}. This paper also reviewed scaling laws for inter-execution times of (state-dependent) homogeneous and polynomial systems.
	
	References \citep{mazo2009self} and also \citep{mazo2010iss} exploited ISS approach for self-triggered control of a linear sampled-data system. The authors have considered a kind of Lyapunov sampling to ensure exponential input-to-state stability of the closed-loop system. From implementation perspective, the authors considered discrete-time version of event-triggering mechanism (similar to periodic event-triggered control techniques). Compared with work in the literature, simulations show that the proposed self-triggered control in this work provides a system with greater sampling intervals as well as explicit guarantees of performance.
	
	In \citep{wang2008state} and \citep{wang2009self}, Wang and Lemmon derived used an self-triggering mechanism  to guarantee finite-gain $ L_2 $ stability of the sampled-data system which required a continuous decay of the Lyapunov function. The authors derived bounds on task's sampling periods and deadlines and some sufficient conditions for the existence of admissible sequences of release and finishing times. It should be noted that the plant is controlled by a full-information $ H_\infty $ controller. In contrast with those two papers, the paper \citep{wang2010self}, considers disturbances which are independent of the process model and it can be any $ L_2 $ signal. The paper \citep{lemmon2007self}, considered Earliest-Deadline-First (EDF) and Buttazzo's elastic scheduling algorithms for scheduling of multiple self-triggered control tasks. 

The main contribution of this paper is introducing optimal self-triggering mechanisms to calculate, at each sampling instant, the next sampling horizon to maximize (under certain criterion) the average sampling interval while guaranteeing the system's exponential stability. The presented mechanisms are developed for unperturbed and perturbed LTI systems.

A main drawback of such self-triggering mechanisms is their computation load. In other words, at each sampling instant, a heavy computational algorithm should be performed to find the solution of an optimization problem. Computationally speaking, it is not applicable in a real-time implementation. Therefore, we will present two versions of the self-triggering mechanism: an online version, to illustrate the approach, where heavy computations are required to be performed during the online control of the system, and an offline, tractable version of the algorithm, which can be implemented on real systems.

The organization of this paper is structured to provide a methodical exploration of the proposed self-triggering mechanism for linear sampled-data systems. In Section \ref{Des}, we detail the system description and problem statement, laying the foundation for the subsequent analysis. Section \ref{Unperturbed} is devoted to presenting the main results for the perturbation-free case, including a comprehensive examination of the proposed self-triggering mechanism and stability analysis for both online and offline scenarios. The discussion continues in Section \ref{Perturbed}, which extends the analysis to the perturbed case, again comparing the online and offline versions of the self-triggering mechanism. Simulation validation is provided in Section \ref{Simulation}, where the results for both perturbation-free and perturbed scenarios are presented, showcasing the efficacy of the proposed mechanisms under various conditions. The paper concludes in Section \ref{Conclusion}, summarizing the findings and contributions of this work. Additionally, Section \ref{Appendix} includes an appendix providing detailed proofs to support the theoretical claims made throughout the paper.

\section{System Description and Problem Statement}\label{Des}

\subsection{System Description}

We consider a continuous time LTI control system
\begin{equation} \label{syst}
	\begin{split}
		\dot{x}(t)&=Ax(t)+Bu(t),~\forall t \geq 0, 
	\end{split}
\end{equation}
where $ x \in \mathbb{R}^n $ is the system's state, $ u \in \mathbb{R}^m $ is the system's input and $ A $ and $ B $ are matrices of appropriate dimensions. We consider a sampled-data sate-feedback controller
\begin{equation} \label{statefc}
	u(t)=Kx(t_h),~ \forall t \in [t_h,t_{h+1}),~h\in \mathbb{N},
\end{equation}
in which $ K $ is a given feedback gain matrix such that $ A+BK $ is Hurwitz, and the sampling instants $ t_h,~h \in \mathbb{N} $, satisfy
\begin{equation} \label{samplint}
	T_h=t_{h+1}-t_h>0, ~\forall h\in \mathbb{N},~\text{with} ~ t_0=0.
\end{equation}

The discrete-time model of the system \eqref{syst}  at instants $ t_h $ can be written as 
\begin{equation} \label{dsyst}
	\begin{split}
		{x}(t_{h+1})&={\tilde{A}}_{(T_h)} x(t_h)
	\end{split}
\end{equation}
with
\begin{equation} \label{dsyst1}
	\begin{split}
		{\tilde{A}}_{(T_h)}=e^{AT_h}+\int\nolimits_{0}^{T_h}e^{As}BK ds
	\end{split}
\end{equation}
The aim of the paper is to propose a self triggering mechanism in which the next sampling instant $ t_{h+1},~ h\in\mathbb{N} $, is calculated at each sampling instant $ t_h,~h\in \mathbb{N} $, based on $ x(t_h) $ and an optimization criterion.

\subsection{Motivational Example}

Consider a system of the form \eqref{dsyst} with a constant sampling interval $ T_h=T,~ \forall h \in \mathbb{N} $, and the following characteristics
\begin{equation}
	A=\begin{bmatrix} 0 & 1\\-2 & 0.1 \end{bmatrix},~ B=\begin{bmatrix} 0 \\1 \end{bmatrix}, ~ K = \begin{bmatrix} 1 & 0 \end{bmatrix}.
\end{equation}

\begin{figure}[h] 
	\centering
	\subfloat{%
		\resizebox*{7cm}{!}{\includegraphics{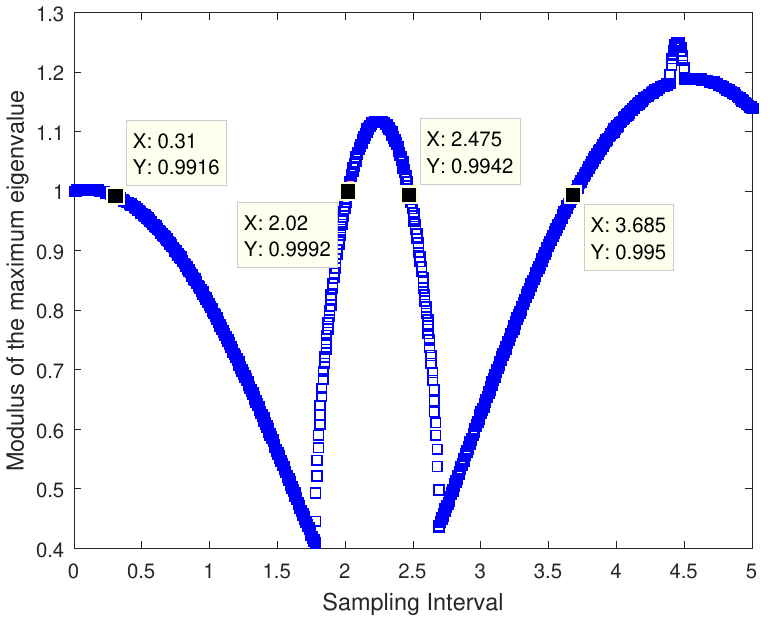}}}\hspace{2pt}
	\caption{ Evolution of the modulus of the maximum eigenvalue $ {\tilde{A}}_{(T_h)} $ with respect to $ T_h $} 	\label{MeigSys}
\end{figure}

Fig.~{ \ref{MeigSys}} shows that the system stability depends on the sampling interval $ T $. The system \eqref{dsyst} with a constant sampling period $ T $ is stable if $ {\tilde{A}}_{(T)} $ is a Schur matrix (i.e. $ \max\Big(\Big|\lambda({\tilde{A}}_{(T)})\Big|\Big)<1 $, where $\lambda(Q) $ refers to the eigenvalues of the matrix $ Q \in \mathbb{R}^{n\times n} $) and we call the sampling interval $ T $ a stabilizing sampling interval. Otherwise, it is called a non stabilizing sampling interval. In the case of non-constant sampling intervals, stability analysis become even more complicated.

To clarify our motivation, let us look at three cases in which combination of stabilizing  and non stabilizing sampling intervals may result in stable or unstable systems.
\begin{itemize}
	\item Case $ 1 $: Consider stabilizing sampling intervals $ {T_s}_1=1.5s $, $ {T_s}_2=3s $, the sampled-data system \eqref{dsyst} is stable with both constant sampling intervals $ {T_s}_1 $ and $ {T_s}_2$. However, the product matrix $ {\tilde{A}}_{({T_s}_1)}{\tilde{A}}_{({T_{s}}_2)} $ is not a Schur matrix, then the sampling sequence $ ({T_s}_1,{T_s}_2,{T_s}_1,{T_s}_2,{T_s}_1,{T_s}_2, \cdots) $ does not stabilize the system, as it is shown in Fig.~{ \ref{case1}}.  	
	\begin{figure}[h] 
		\centering
		\subfloat[Constant Sampling $ {T_s}_1=1.5s$-Stable]{\resizebox*{4.3cm}{!}{\includegraphics{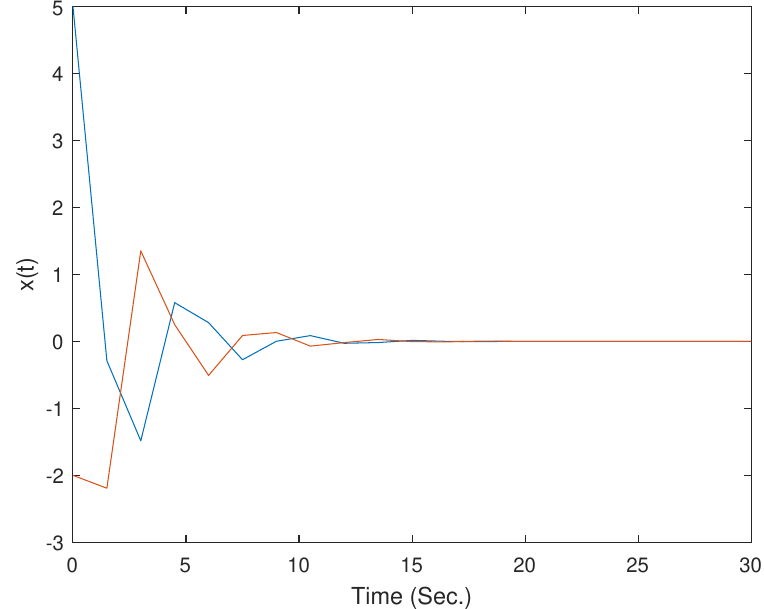}}}
		\subfloat[Constant Sampling $ {T_s}_2=3s $-Stable]{%
			\resizebox*{4.3cm}{!}{\includegraphics{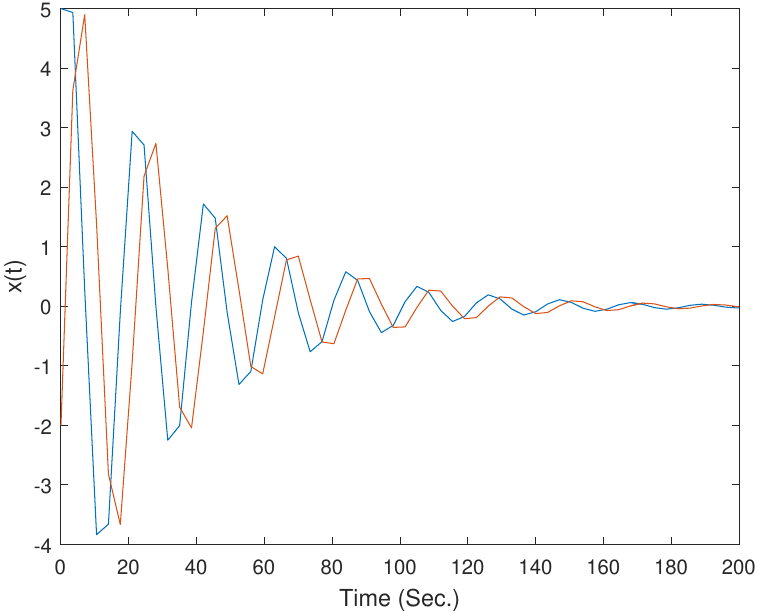}}}
		\subfloat[Variable Sampling $ {T_s}_1 \rightarrow {T_s}_2 \rightarrow {T_s}_1 \rightarrow {T_s}_2 \rightarrow \cdots $-Unstable]{\resizebox*{4.3cm}{!}{\includegraphics{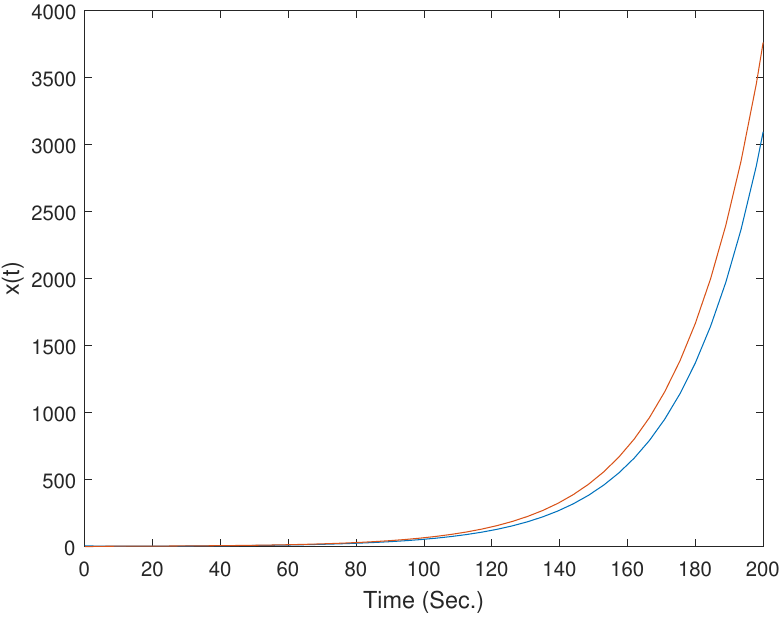}}}
		\caption{ Combination of two stabilizing sampling interval results in unstable system } 	\label{case1}
	\end{figure}	
	\item Case $ 2 $: Consider non stabilizing sampling intervals $ {T_{ns}}_1=2.126s $, $ {T_{ns}}_2=3.95s $, the sampled-data system \eqref{dsyst} is not stable with both constant sampling intervals $ {T_{ns}}_1 $ and $ {T_{ns}}_2$. However, the product matrix $ {\tilde{A}}_{{(T_{ns})}_1}{\tilde{A}}_{{(T_{ns})}_2} $ is a Schur matrix, then  the sampling sequence $ ({T_{ns}}_1,{T_{ns}}_2,{T_{ns}}_1,{T_{ns}}_2,{T_{ns}}_1,{T_{ns}}_2, \cdots) $ stabilizes the system and it is asymptotically stable, as it is shown in Fig.~{ \ref{case2}}.	
	\begin{figure}[h] 
		\centering
		\subfloat[Constant Sampling $ {T_{ns}}_1=2.126s $-Unstable]{%
			\resizebox*{4.3cm}{!}{\includegraphics{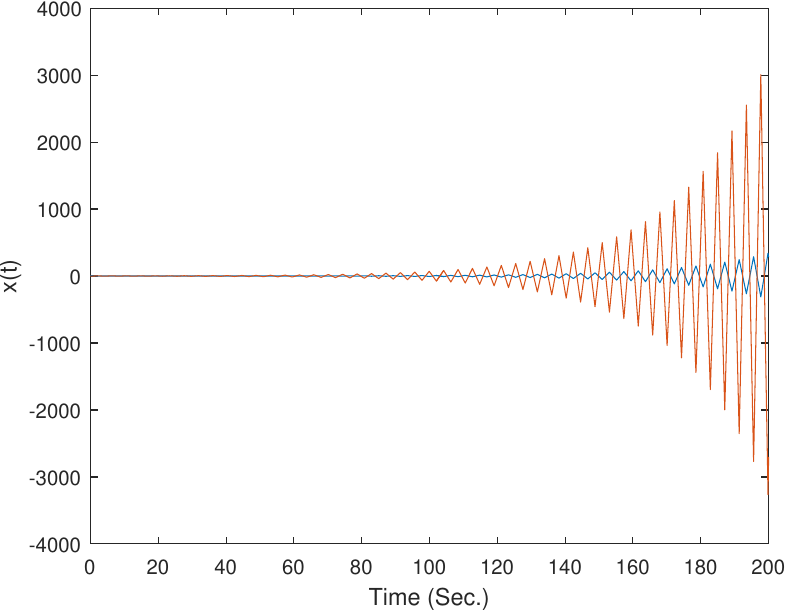}}}
		\subfloat[Constant Sampling $ {T_{ns}}_2=3.95s $-Unstable]{%
			\resizebox*{4.3cm}{!}{\includegraphics{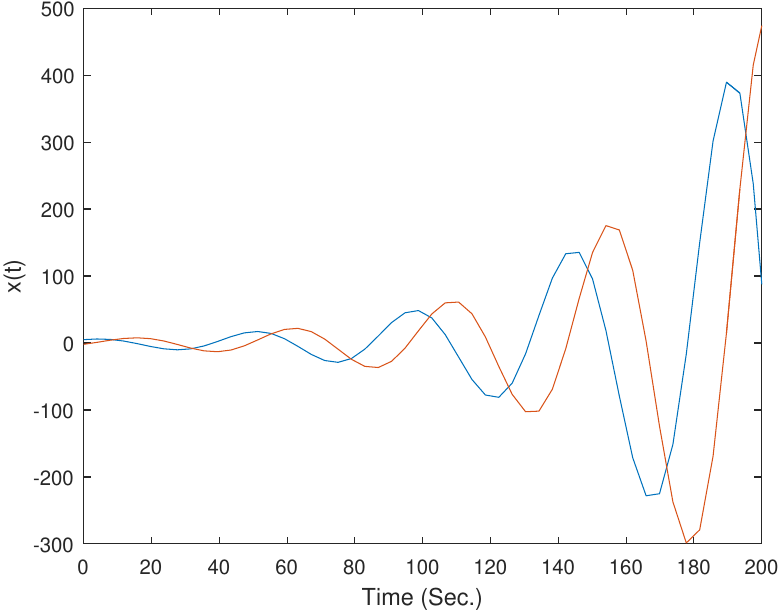}}}
		\subfloat[Variable Sampling $ {T_{ns}}_1 \rightarrow {T_{ns}}_2 \rightarrow {T_{ns}}_1 \rightarrow {T_{ns}}_2 \rightarrow \cdots $-Stable]{%
			\resizebox*{4.3cm}{!}{\includegraphics{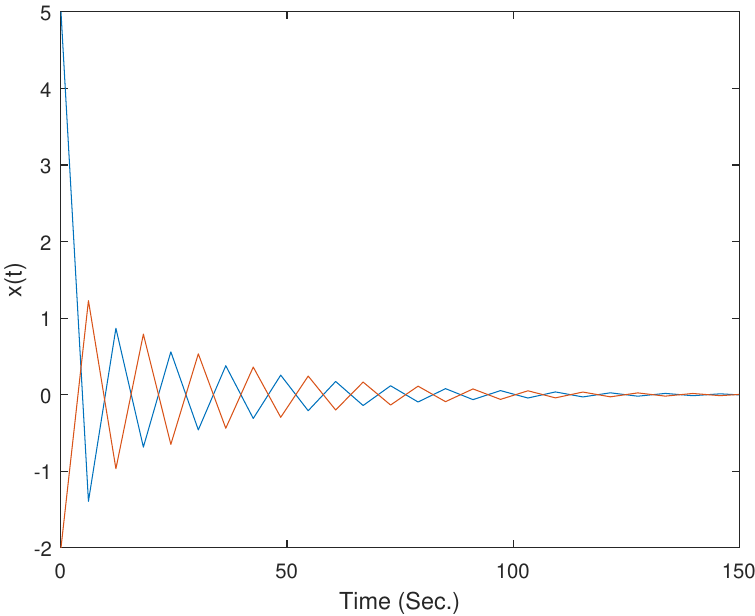}}}\hspace{2pt}
		\caption{ Combination of two non stabilizing sampling interval results in stable system} 	\label{case2}
	\end{figure}	
	\item Case $ 3 $: consider a non stabilizing sampling interval $ T_{ns}=2.126s $ and a stabilizing sampling interval $ T_{s}=2.9s $. Since the product matrix $ {\tilde{A}}_{(T_s)}{\tilde{A}}_{(T_{ns})} $ is a Schur matrix, then  the sampling sequence $ (T_{ns},T_s,T_{ns},T_s,T_{ns},T_s, \cdots) $ stabilizes the system, as it is shown in Fig.~{ \ref{case3}}.	
	\begin{figure}[h] 
		\centering
		\subfloat[Constant Sampling $ {T_{ns}}=2.126s $-Unstable]{%
			\resizebox*{4.3cm}{!}{\includegraphics{T1Unstable-Unstable}}}
		\subfloat[Constant Sampling $ {T_{s}}=2.9s $-Stable]{%
			\resizebox*{4.3cm}{!}{\includegraphics{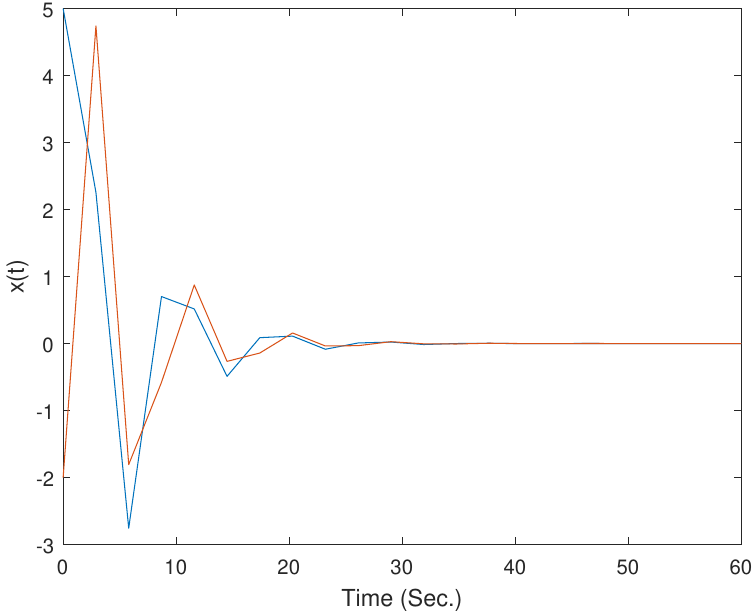}}}
		\subfloat[Variable Sampling $ {T_{ns}} \rightarrow {T_{s}} \rightarrow {T_{ns}} \rightarrow {T_{s}} \rightarrow \cdots $-Stable]{%
			\resizebox*{4.3cm}{!}{\includegraphics{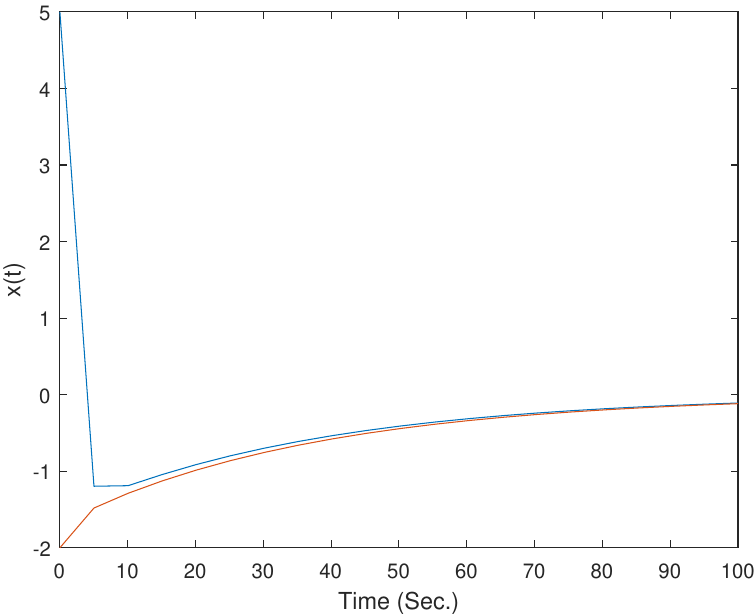}}}\hspace{2pt}
		\caption{ Combination of a non stabilizing sampling interval with a stabilizing sampling interval results in stable system} 	\label{case3}
	\end{figure}	
\end{itemize}

Motivated by this observation, this work aims at characterizing other sampling sequences that not only stabilize the system \eqref{dsyst} but also have a maximum average of sampling intervals among other possible sequences based on the current state of the system. 

\subsection{System Reformulation}\label{sys-refor}

In classical work concerning self triggering control, at each sampling instant, the next optimal (based on some certain criteria) admissible sampling intervals are computed. In this work, we want to compute the next optimal sampling intervals over a finite horizon. Therefore, we need to introduce a few notations regarding sampling horizons and sampling sequences.

$ \sigma=\{T_{\sigma}^j\}_{j=1}^{l}=(T_{\sigma}^1,\cdots,T_{\sigma}^l) $, with $l \in \mathbb{N} \setminus \{0\} $, refers to a sampling horizon of length $ l $, where $ j $ represents the position of a sampling interval $ T_{\sigma}^j $ inside the horizon.  For a finite set $ \Gamma \subset \mathbb{R}^+ $, we define $ S_{l_{min}}^{l_{max}}(\Gamma) $ the set of all horizons $ \sigma=\{T_{\sigma}^j \}_{j=1}^l $ of length $ l \in [l_{min},l_{max}] $ with values $ T_{\sigma}^j \in \Gamma,~ \forall j\in \{1,\cdots,l\} $:
\begin{equation}\label{key}
	\begin{split}
	S_{l_{min}}^{l_{max}}(\Gamma) =\Big \{ \sigma=\{T_{\sigma}^j\}_{j=1}^{l}: l \in [l_{min},l_{max}],~\text{and}\\
	T_{\sigma}^j \in \Gamma ,~\forall j\in \{1,\cdots,l\} \Big \}.
	\end{split}
\end{equation}

\begin{figure}[h] 
	\centering
	\subfloat{%
		\resizebox*{8cm}{!}{\includegraphics{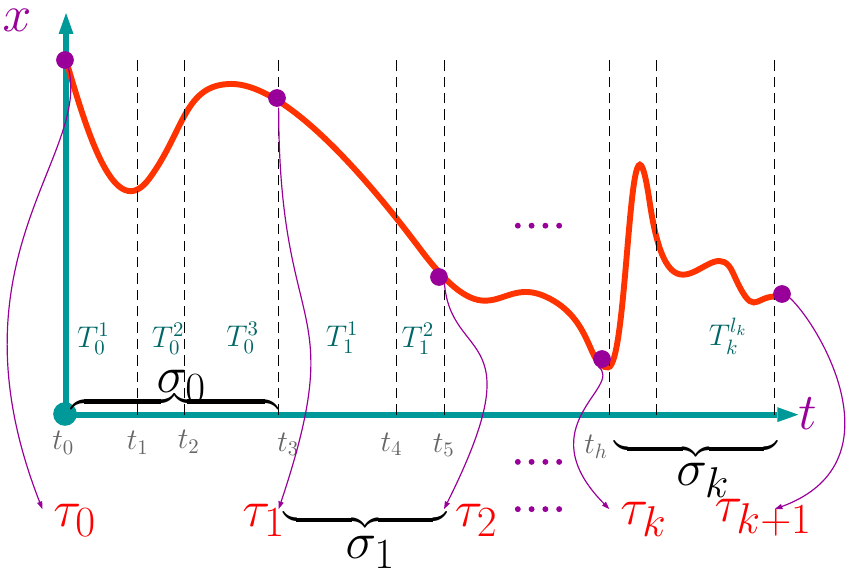}}}\hspace{2pt}
	\caption{ System state discretization using variable sequences } 	\label{Instant}
\end{figure}

We denote by $ \{\sigma_k\}_{k\in \mathbb{N}}$ a sampling sequence composed by sampling horizons $ \sigma_k=(T_{\sigma_k}^1,\cdots,T_{\sigma_k}^{l_k}) \in S_{l_{min}}^{l_{max}}(\Gamma),~ k \in \mathbb{N}  $. The scalars $ T_{\sigma_k}^{i}, ~ k \in \mathbb{N},~ i\in \{1,\cdots,l_k\} $ define sampling intervals which constitute the sampling sequence where
$ k $ indicates the index of the horizon and $ i $ the position of this sampling step in the considered horizon $ \sigma_k $. 
The representation of system \eqref{dsyst} over a sampling sequence $\{\sigma_k\}_{k\in \mathbb{N}} $ is given as
\begin{equation} \label{Hdsyst}
	\begin{split}
		{x}{_{k+1}}=\Phi_{\sigma_k} x_k ,~ k\in \mathbb{N},
	\end{split}
\end{equation}
with
\begin{equation} \label{Hdsyst1}
	\begin{split}
		\Phi_{\sigma_k}={\tilde{A}}_{(T^{l_k}_{\sigma_k})}{\tilde{A}}_{(T^{l_{k-1}}_{\sigma_k})}\cdots {\tilde{A}}_{(T^{l_{1}}_{\sigma_k})},~x_k=x(\tau_k).
	\end{split}
\end{equation}
The instant $ \tau_k,~ {k\in \mathbb{N}} $ denotes the starting time of a sampling sequence $ \sigma_k $ such that

\begin{equation}
	\tau_{k+1}=\tau_{k}+\sum\nolimits_{i=1}^{l_{k}}T_{\sigma_k}^i,~ \tau_0=t_0=0  
\end{equation} 
A description of these sequences is depicted in Fig.~{ \ref{Instant}}. 


\subsection{Problem Statement}

Our aim is to calculate, for a current state $ x_k=x(\tau_k) $ at instant $ \tau_k $, the next sampling horizon $ \sigma_k \in S_{l_{min}}^{l_{max}}(\Gamma) $ for a given $ l_{min}$ and  $l_{max} $ and $ \Gamma $ which has a maximum possible average sampling interval and  ensures the system's exponential stability.

\section{Main Results: Perturbation-free Case}\label{Unperturbed}
\subsection{Proposed Self Triggering Mechanism and Stability Analysis: Online case}\label{STM-off}

\subsubsection{Self Triggering Mechanism and Stability Analysis}
Proposition \ref{S_O_Df} ensures exponential stability of the discretized system \eqref{Hdsyst} in which the next sampling horizon $ \sigma_k $ defined by proposed self-triggering mechanism. 
\begin{proposition}\label{S_O_Df}
	Consider a scalar $ \beta>0 $ and a matrix $ P=P^T\succ0 $ such that
	\begin{equation} \label{matrixp}
		\begin{split}
			&\exists \sigma^* \in S^{l_{max}}_{l_{min}}(\Gamma): \Phi^T_{\sigma^*} P \Phi_{\sigma^*} -e^{(-\beta\sum\nolimits_{j=1}^{|{\sigma^*} |}T_{\sigma^*}^{j }) }P\prec0
		\end{split} .
	\end{equation}
	
	Then, system \eqref{Hdsyst} with the self-triggering mechanism defined by
	
	\begin{equation} \label{STM-on-wp}
		\begin{split}
			& \sigma_k  \in \argmax_{{\sigma} \in {{\bar{S}}^{l_{max}}_{l_{min}}}(\Gamma,x_k)}\Bigg\{ \frac{\sum\nolimits_{j=1}^{|{\sigma} |}T^j_{{\sigma} }}{|{\sigma} |} \Bigg\},
		\end{split}
	\end{equation}	
	with
	\begin{equation}\label{es-wp}
		\begin{split}
		&{{\bar{S}}^{l_{max}}_{l_{min}}}(\Gamma,x_k) = \\
		&\Bigg\{\sigma \in S^{l_{max}}_{l_{min}}(\Gamma): x^T_k \Big(\Phi^T_{\sigma} P \Phi_{\sigma} -e^{(-\beta\sum\nolimits_{j=1}^{|{\sigma} |}T_{\sigma}^j) }P  \Big)x_k \leq 0 \Bigg\},
	\end{split}
	\end{equation}
	is exponentially stable with the decay rate of $ \beta/2 $.
\end{proposition}

\begin{proof}
	Consider a quadratic Lyapunov function $ V:\mathbb{R}^n  \to \mathbb{R}^+ $ defined by $ V(x_k)=x^T_k P x_k $ in which the matrix $ P $ satisfies \eqref{matrixp}. Then by construction, the set $ {{\bar{S}}^{l_{max}}_{l_{min}}}(\Gamma,x_k) $ defined by \eqref{es-wp} is not empty (i.e. $ \sigma^* \in {\bar{S}}^{l_{max}}_{l_{min}}(\Gamma,x_k),~ \forall x_k \in \mathbb{R}^n $). Therefore,  the arguments of the maxima in the equation \eqref{STM-on-wp} is well-defined and non-empty. Then $ \forall k \in \mathbb{N} $, by the equation \eqref{STM-on-wp} we can conclude that
	\begin{equation}
		\begin{split}
			V(x_{k+1})&= x_{k+1}^T P x_k = x^T_k \Phi^T_{{\sigma}_k} P\Phi_{{\sigma}_k} x_k \\
			& \leq  e^{(-\beta\sum\nolimits_{j=1}^{|{\sigma_k} |}T_{\sigma_k}^{j }) }x^T_k P x_k = e^{-\beta (\tau_{k+1}-\tau_k)} V(x_k).
		\end{split}
	\end{equation}
	Therefore, $ \forall k \in \mathbb{N} $, by recursion, we have  $ V(x_k) \leq e^{-\beta \tau_k} V(x_0) $, and according to \cite{khalil2002nonlinear}, system \eqref{Hdsyst} with self triggering control \eqref{STM-on-wp} is exponentially stable with the decay rate of $ \beta/2 $.
	
\end{proof}

\begin{remark}
	Self-triggering mechanism defined by \eqref{STM-on-wp} chooses a next optimal sequence from the set \eqref{es-wp} which contains all stable sequences such that the chosen sequence has a maximum average of sampling intervals among others.
\end{remark}

\subsubsection{Implementation Algorithm }
Algorithm \ref{Alg1} provides a self triggering mechanism for the sampled-data system \eqref{Hdsyst} which generates the next stabilizing and optimal sampling horizon $ {\sigma}_k $ based on the sampled state  $ x_k=x(\tau_k) $.

	\begin{algorithm}[!htbp]
		\caption{Self Triggering Mechanism for Sampled-Data System: Online case}\label{Alg1}
		\begin{algorithmic}
			\STATE
			\STATE \textbf{\underline{OFFLINE Procedure:}}
			\begin{itemize}
				\STATE Define $ l_{min} $ and $ l_{max} $, the minimum and maximum length of the considered sampling horizons.
				\STATE Define a finite set $ \Gamma=\{T^1,\cdots,T^m\} $ of sampling intervals.
				\STATE Design the set $ S^{l_{max}}_{l_{min}}(\Gamma) $ of sampling horizons of length $l \in [l_{min},l_{max}] $ constituted of sampling intervals in $ \Gamma $.
				\STATE Compute the matrix $ P=P^T\succ0 $ for a given stable sequence $ \sigma^*\in S^{l_{max}}_{l_{min}}(\Gamma) $ i.e. $ \Phi_{\sigma^*} $ is a Schur matrix satisfying: $ \Phi^T_{\sigma^*} P \Phi_{\sigma^*} -e^{(-\beta\sum\nolimits_{j=1}^{|{\sigma^*} |}T_{\sigma^*}^{j }) }P\prec0$.
			\end{itemize}
			\STATE
			\STATE \textbf{\underline{ONLINE Procedure:} At each instant $ \tau_k $:}
			\STATE $ {T_{avg}}_{max}=\frac{\sum\nolimits_{j=1}^{|{\sigma^*} |}T_{\sigma^*}^{j }}{|{\sigma^*} |} $
			\STATE $ {\sigma_{opt}}=\{\sigma^*\} $
			\FORALL{$ \sigma \subset S^{l_{max}}_{l_{min}}(\Gamma) $}
			\IF{$ x^T_k\Big(\Phi^T_{\sigma} P \Phi_{\sigma} -e^{(-\beta\sum\nolimits_{j=1}^{|{\sigma} |}T_{\sigma}^{j }) }P\Big)x_k < 0 $} 
			\STATE $ T_{avg}=\frac{\sum\nolimits_{j=1}^{|{\sigma} |}T_{\sigma}^{ j}}{|{\sigma} |}$ 
			\IF {$ T_{avg} >{T_{avg}}_{max}$} 
			\STATE $ {T_{avg}}_{max}= T_{avg}$
			\STATE $ {\sigma_{opt}}=\{\sigma\} $
			\ELSIF {$ T_{avg} ={T_{avg}}_{max}$}
			\STATE $ {\sigma_{opt}}=\{\sigma_{opt},\sigma\} $       
			\ENDIF        
			\ENDIF            
			\ENDFOR
			\STATE Choose randomly  $ \sigma_k$, from $ {\sigma_{opt}} $.    
		\end{algorithmic}
	\end{algorithm}

\subsection{Proposed Self Triggering Mechanism and Stability Analysis: Offline case}\label{wp-online}
A main drawback of the algorithm which is introduced in the previous is its heavy computation load. In other words, at each instant $ \tau_k $, Algorithm \ref{Alg1} should be performed to find the solution of the optimization problem. Computationally speaking, it is not applicable in a real-time implementation. For example, Algorithm \ref{Alg1} needs to examine the set of all horizons to find stable sequences and accordingly all optimal sequences. Thus, the algorithm has a complexity  of $ \mathcal{O}\Big\{\Big|S_{l_{min}}^{l_{max}}(\Gamma)\Big|=\sum_{i=l_{min}}^{l_{max}} \Big|\Gamma\Big|^i\Big\} $, which is not realistic for a real-time implementation.

In this section, an offline tractable method for partitioning of the state space into a finite number of conic regions $ \mathcal{R}_c,~c \in \{1,\cdots,N\} $, is proposed. The idea is to define a set $ \Psi_{(c)} $ of optimal sequences for each of these regions. For this purpose, we use conic covering technique proposed in \cite{fiter2012state}:

Consider  $ N \in \mathbb{N}$ conic regions $ \mathcal{R}_c$ such that
\begin{equation}\label{conicreg}
	\mathcal{R}_c=\Big\{x \in \mathbb{R}^n: x^T Q_c x \geq 0 \Big\} ,~\forall c \in \{1,\cdots,N\},
\end{equation}
where the matrices $ Q_c $ are designed as in \cite{fiter2012state}.

\begin{figure}[h] 
	\centering
	\subfloat{%
		\resizebox*{8cm}{!}{\includegraphics{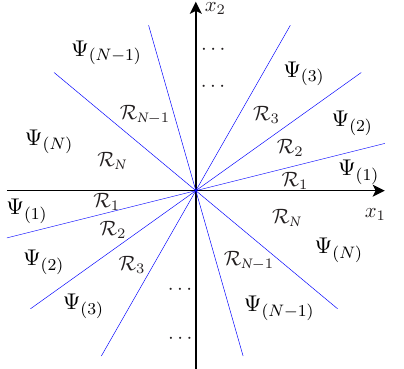}}}\hspace{2pt}
	\caption{ Covering the state space of dimension 2 with $ N $ conic regions $ \mathcal{R}_c, c \in \{1,\cdots,N\}  $} 	\label{Region1}
\end{figure}

A partition of the state space based on this technique is shown in Fig.~{ \ref{Region1}}. 

\subsubsection{Self Triggering Mechanism and Stability Analysis}
Proposition \ref{STM_on} ensures exponential stability of the discretized system \eqref{Hdsyst} in which for a conic region $ \mathcal{R}_c $, the next sampling horizon $ \sigma^{(c)} $ defined by proposed self-triggering for each conic region $ \mathcal{R}_c,~c \in \{1,\cdots,N\} $. 

\begin{proposition}\label{STM_on}
	Consider a scalar $ \beta>0 $ and a matrix $ P=P^T\succ0 $ such that	
	\begin{equation} \label{matrixp-off}
		\begin{split}
			& \exists \sigma^* \in S^{l_{max}}_{l_{min}}(\Gamma): \Phi^T_{\sigma^*} P \Phi_{\sigma^*} -e^{(-\beta\sum\nolimits_{j=1}^{|{\sigma^*} |}T_{\sigma^*}^{j }) }P\prec0.
		\end{split} 
	\end{equation}
	
	Then the system \eqref{Hdsyst} with the self-triggering mechanism defined by:	
	\begin{equation} \label{STM-off-wp}
		\begin{split}
			& \sigma_{k}  \in \argmax_{\sigma \in \Psi_{(c)}:~ c \in \{1,\cdots,N\}~\text{and}~ x_k \in \mathcal{R}_c} \Bigg\{ \frac{\sum\nolimits_{j=1}^{|{\sigma} |}T_{\sigma}^{j }}{|{\sigma} |}  \Bigg\},
		\end{split}
	\end{equation}
	with
	\begin{equation} 
		\begin{split}
			& \Psi_{(c)}  = \argmax_{{\sigma} \in {{\bar{S}}^{l_{max}}_{l_{min}}}(\Gamma,\mathcal{R}_c)}\Bigg\{ \frac{\sum\nolimits_{j=1}^{|{\sigma} |}T_{\sigma}^{j }}{|{\sigma} |}  \Bigg\},
		\end{split}
	\end{equation}
	where	
	\begin{equation}\label{es-off-wp}
	\begin{split}
		{{\bar{S}}^{l_{max}}_{l_{min}}}(\Gamma,\mathcal{R}_c) =\bigg \{\sigma \in S^{l_{max}}_{l_{min}}(\Gamma): \exists\epsilon_c>0 ,\\
		\Phi^T_{\sigma} P \Phi_{\sigma} -e^{(-\beta\sum\nolimits_{j=1}^{|{\sigma} |}T_{\sigma}^{j }) }P+\epsilon_c Q_c   \preceq 0 \bigg\},
	\end{split}
	\end{equation}
	is exponentially stable with the decay rate of $ \beta/2 $.
\end{proposition}

\begin{proof}
	Consider a quadratic Lyapunov function $ V:\mathbb{R}^n  \to \mathbb{R}^+ $ defined by $ V(x_k)=x^T_k P x_k $ in which the matrix $ P $ satisfies \eqref{matrixp-off}. Then by construction, the set $ {{\bar{S}}^{l_{max}}_{l_{min}}}(\Gamma,\mathcal{R}_c) $ defined by \eqref{es-off-wp} is not empty (i.e. $ \sigma^* \in {\bar{S}}^{l_{max}}_{l_{min}}(\Gamma,x_k),~ \forall x_k \in \mathbb{R}^n $). Therefore,  the arguments of the maxima in the equation \eqref{STM-off-wp} is well-defined and non-empty. Let $ x_k \in \mathbb{R}^n, k\in \mathbb{N} $, there exists a conic region $ \mathcal{R}_c $ as in \eqref{conicreg} such that $ x_k \in  \mathcal{R}_c$. Consider the sampling horizon $ \sigma_k $ as defined in \eqref{STM-off-wp}. Therefore, we have $ x^T_{k} \Big(\Phi^T_{\sigma_k} P \Phi_{\sigma_k} -e^{(-\beta\sum\nolimits_{j=1}^{|{\sigma_k} |}T_{\sigma_k}^j) }P  \Big)x_{k} \leq x^T_{k} \Big(\Phi^T_{\sigma_k} P \Phi_{\sigma_k} -e^{(-\beta\sum\nolimits_{j=1}^{|{\sigma_k} |}T_{\sigma_k}^j) }P +\epsilon_c Q_c \Big)x_{k} \leq 0$, since by construction $ x_k\in \mathcal{R}_c $ and $ \sigma_k \in {{\bar{S}}^{l_{max}}_{l_{min}}}(\Gamma,\mathcal{R}_c) $. Then, one can conclude that
	
	\begin{equation}
		\begin{split}
			V(x_{k+1})&=x^T_{k+1}Px^T_{k+1}=x^T_k \Phi^T_{{\sigma}_k} P\Phi_{{\sigma}_k} x_k \\
			& \leq  e^{(-\beta\sum\nolimits_{j=1}^{|{\sigma} |}T_{\sigma}^{j }) }x^T_k P x_k= e^{-\beta (\tau_{k+1}-\tau_{k})} V(x_k).
		\end{split}
	\end{equation}
	Therefore, $ \forall k\in\mathbb{N} $, by recursion, we have $ V(x_k) \leq e^{-\beta \tau_k} V(x_0) $ and according to \cite{khalil2002nonlinear}, system \eqref{Hdsyst} with self triggering control \eqref{STM-off-wp} is exponentially stable with the decay rate of $ \beta/2 $.  
\end{proof}

\begin{remark}
	This should be noted that in this theorem, the larger the number of regions by which the state space is partitioned, the greater the performance the system will have.
\end{remark}

\subsubsection{Implementation Algorithm}
Algorithm \ref{Alg2} provides a self triggering algorithm for the sampled-date system \eqref{Hdsyst} which guarantees exponential stability and generates the next stabilizing optimal sampling horizon $ {\sigma} \in \Psi_{(c)} $ based on $ x(\tau_k) \in \mathcal{R}_c $ such that the selected sampling horizon has a maximum average of the sampling intervals:

	\begin{algorithm}[!htbp]
		\caption{Self Triggering Mechanism for Sampled-Data System: Offline case}\label{Alg2}
		\begin{algorithmic}
			\STATE
			\STATE \textbf{\underline{OFFLINE Procedure:}}
			\begin{itemize}
				\STATE Define $ N $, the number of conic regions.
				\STATE Define $ l_{min} $ and $ l_{max} $, the minimum and maximum length of the considered sampling horizons.
				\STATE Define a finite set $ \Gamma=\{T^1,\cdots,T^m\} $ of sampling intervals.
				\STATE Design the set $ S^{l_{max}}_{l_{min}}(\Gamma) $ of sampling horizons of length $l \in [l_{min},l_{max}] $ constituted of sampling intervals in $ \Gamma $.
				\STATE Compute the matrix $ P=P^T\succ0 $ for a given stable sequence $ \sigma^*\in S^{l_{max}}_{l_{min}}(\Gamma) $ i.e. $ \Phi_{\sigma^*} $ is a Schur matrix satisfying: $ \Phi^T_{\sigma^*} P \Phi_{\sigma^*} -e^{(-\beta\sum\nolimits_{j=1}^{|{\sigma^*} |}T_{\sigma^*}^{j }) }P\prec0$.
				\STATE Design a matrix $ Q_c $ for each conic region $ \mathcal{R}_c=\{x \in \mathbb{R}^n: x^T Q_c x \geq 0 \},~\forall c \in \{1,\cdots,N\} $.
				\STATE Compute the set $ \Psi_{(c)} $ of optimal sampling horizons for each conic region $ \mathcal{R}_c $:
				\begin{itemize}
					\FORALL{$c=1:N$} 
					\STATE $ {T_{avg}}_{max}=\frac{\sum\nolimits_{j=1}^{|{\sigma^*} |}T_{\sigma^*}^{j }}{|{\sigma^*} |} $
					\STATE $ {\Psi^{opt}_{(c)}}=\{\sigma^*\} $
					\FORALL{$ \sigma \subset S^{l_{max}}_{l_{min}}(\Gamma) $}
					\IF{$ \Phi^T_{\sigma} P \Phi_{\sigma} -e^{(-\beta\sum\nolimits_{j=1}^{|{\sigma} |}T_{\sigma}^{j }) }P+\epsilon_c Q_c\prec0 $}
					\STATE $ T_{avg}=\frac{\sum\nolimits_{j=1}^{|{\sigma} |}T_{\sigma}^{j }}{|{\sigma} |}$
					\IF{$ T_{avg} > {T_{avg}}_{max}$}
					\STATE $ {T_{avg}}_{max}= T_{avg}$
					\STATE $ {\Psi^{opt}_{(c)}}=\{\sigma\} $
					\ELSIF{$ T_{avg} ={T_{avg}}_{max}$}
					\STATE $ {\Psi^{opt}_{(c)}}=\{\Psi^{opt}_{(c)},\sigma\} $
					\ENDIF
					\ENDIF
					\ENDFOR
					\STATE $ \Psi_{(c)}={\Psi^{opt}_{(c)}} $
					\ENDFOR
				\end{itemize}
			\end{itemize}
			\STATE
			\STATE \textbf{\underline{ONLINE Procedure:} At each instant $ \tau_k $:}
			\begin{itemize}
				\STATE Determine the conic region $ \mathcal{R}_c,~c \in \{1,\cdots,N\} $, in which current state $ x_k $ belongs.
				\STATE Select randomly a corresponding optimal horizon $ \sigma_k $ from the sets $ \Psi_{(c)},~c \in \{1,\cdots,N\} $, for which $ x_k \in \mathcal{R}_c $.
			\end{itemize}
		\end{algorithmic}
	\end{algorithm}

\section{Main Results: Perturbed Case}\label{Perturbed}
In this section, a self-triggered mechanism is introduced which maximizes the average of next sampling intervals regarding given set of user-defined inter-sampling periods for a continouse time LTI system under bounded perturbation. 
\subsection{System Reformulation}\label{sysref}
We consider a perturbed continuous time LTI system
\begin{equation} \label{systP}
	\begin{split}
		\dot{x}(t)=Ax(t)+Bu(t)+Dw(t),~ t\geq 0,\\
		\quad x \in \mathbb{R}^n, ~ u \in \mathbb{R}^m, ~ w \in \mathbb{R}^{n_w}, 
	\end{split}
\end{equation}
where $ w(t) $ is a bounded perturbation. As before, we suppose that the state $ x(t) $ of the system \eqref{systP} is sampled at instants $ t_h $ with $ h \in \mathbb{N}  $ and the sampled-date sate-feedback controller satisfies \eqref{statefc} and \eqref{samplint}. Then, the discrete model of the system at sampling instants $ t_h,~h \in \mathbb{N} $ is given by
\begin{equation}\label{AdsystP}
	\begin{split}
		x(t_h+T_h)=A_{(T_h)}x(t_h)+B_{(T_h)} K x(t_h)+\tilde{w}_{(T_h)}(t_h),
	\end{split}
\end{equation}
with 
\begin{equation}\label{dist_disc}
	\begin{split}
	A_{(T_h)}= e^{AT_h},~ B_{(T_h)} =\int\nolimits_{0}^{T_h}e^{As}Bds,~ \tilde{w}_{(T_h)}(t_h)=\\
	  \int\nolimits_{0}^{T_h}e^{As}Dw(t_h+s) ds.
	\end{split}
\end{equation}

\begin{assumption}\label{pertbounded}
	We suppose that the perturbation is bounded: 
	
	\begin{equation}
		\Big\| \tilde{w}_{(T_h)}(t_h)\Big\|_2 \leq \varpi,~ h\in \mathbb{N}.
	\end{equation} 
\end{assumption}

According to the notation used in the section \ref{sys-refor}, after $ l_h \geq 1 $ steps, one can obtain,
\begin{equation}
	\begin{split}
		x(t_h+T_h^1) &={A}_{(T_h^1)}x(t_h)+{B}_{(T_h^1)} K x(t_h)+\tilde{w}_{(T_h^1)}(t_h)\\
		x(t_h+T_h^1+T_h^2) &={A}_{(T_h^2)}x(t_h+T_h^1)+{B}_{(T_h^2)} K x(t_h+T_h^1) \\
		&\quad +\tilde{w}_{(T_h^2)}(t_h+T_h^1)\\
		&=\bigg({A}_{(T_h^2)}+{B}_{(T_h^2)} K\bigg)  \\
		&\quad \times \bigg[\big({A}_{(T_h^1)}+{B}_{(T_h^1)} K\big)x(t_h)+\tilde{w}_{(T_h^1)}(t_h)\bigg]\\
		&\quad +\tilde{w}_{(T_h^2)}(t_h+T_h^1)\\
		&={\tilde{A}}_{(T_h^2)}{\tilde{A}}_{(T_h^1)}x(t_h)+{\tilde{A}}_{(T_h^2)}\tilde{w}_{(T_h^1)}(t_h) \\
		& \quad +\tilde{w}_{(T_h^2)}(t_h+T_h^1)\\
		&\vdots\\
		x(t_h+\sum_{s=1}^{l_h}T_h^{s})	&=\Bigg(\prod_{s=1}^{l_h}{\tilde{A}}_{(T_h^s)} \Bigg) x(t_h)\\
		& \quad +\Bigg(\sum_{i=0}^{l_h-1}\Bigg(\prod_{j=i+2}^{l_h}{\tilde{A}}_{(T_h^j)}\Bigg)\tilde{w}_{(T_h^{i+1})}\\
		&\quad \times \Big(t_h+\sum_{f=1}^{i}T_h^f\Big)\Bigg)
	\end{split}
\end{equation}
where $ {\tilde{A}}_{(T_h^s)}= {A}_{(T_h^s)}+{B}_{(T_h^s)} K,~\forall s\in \mathbb{N}$.

The representation of system \eqref{AdsystP} over a sampling horizon  $\sigma_k $ is given by
\begin{equation}\label{CLSP}
	\begin{split}
		{x}{_{k+1}}& =\Phi_{\sigma_k} x_k \\
		&\quad +\Bigg(\sum_{i=0}^{|\sigma_k|-1}\Bigg(\prod_{j=i+2}^{|\sigma_k|}{\tilde{A}}_{(T_{\sigma_k}^j)}\Bigg)\tilde{w}_{(T_{\sigma_k}^{i+1})}\Big(\tau_k+\sum_{f=1}^{i}T_{\sigma_k}^f\Big) \Bigg)\\
		&=\Phi_{\sigma_k} x_k+\bar{w}_k,
	\end{split}
\end{equation}
in which $ x_k=x(\tau_k) $ and the transition matrix corresponding to the sampling horizon $ \sigma_k $, from instant $ \tau_k $ to instant $ \tau_{k+1} $ is given by 
\begin{equation}
	\Phi_{\sigma_k}={\tilde{A}}_{(T^{l_k}_{\sigma_k})}{\tilde{A}}_{(T^{l_{k-1}}_{\sigma_k})}\cdots {\tilde{A}}_{(T^{l_{1}}_{\sigma_k})}
\end{equation} 
where 
\begin{equation}
	\tau_{k+1}=\tau_{k}+\sum\limits_{i=1}^{l_{k}}T^i_{\sigma_k}.
\end{equation} 
Since  $ {\tilde{A}}_{(T_{\sigma_k}^s)},~\forall s,k \in \mathbb{N}  $, is a bounded operator and the set $ \Gamma  $ of considered sampling intervals is finite, there exists a constant $ C $ such that 
\begin{equation}
	C=\max_{T_{\sigma_k}^s \in \Gamma} \Big(\Big\|{\tilde{A}}_{(T_{\sigma_k}^s)}\Big\|_2\Big),~\forall k,s \in \mathbb{N},
\end{equation}
According to Assumption \ref{pertbounded},  one can conclude that 
\begin{equation}
	\begin{split}
	\Big\|\bar{w}_k\Big\|_2= \Bigg\|\sum_{i=0}^{|\sigma_k|-1}\Bigg(\prod_{j=i+2}^{|\sigma_k|}{\tilde{A}}_{(T_{\sigma_k}^j)}\Bigg)\tilde{w}_{(T_{\sigma_k}^{i+1})}\Big(\tau_k+\sum_{f=1}^{i}T_{\sigma_k}^f\Big)\Bigg\|_2 \\
	\leq \varpi \sum_{q=0}^{|\sigma_k|-1}C^q .
	\end{split}
\end{equation}
\begin{definition}\cite{khalil2002nonlinear}
	Consider system 
	\begin{equation}\label{exbound}
		\dot{x}=f(x) 
	\end{equation} 
	where $ f:\mathbb{R}^n \to \mathbb{R}^n $ is locally Lipschitz in $ x $. 
	
	The solution of  \eqref{exbound} is Globally Uniformly Ultimately Bounded (GUUB) with ultimate bound $ b>0 $ independent of $ t_0 \geq 0 $, and for every arbitrary large $ a > 0 $, there is $ T=T(a,b)\geq 0 $, independent of $ t_0 $, such that
	\begin{equation}\label{exbdef}
		\Big\|x(t_0) \Big\| \leq a  \Rightarrow \Big\|x(t)\Big\| \leq b, \quad \forall t \geq t_0+T
	\end{equation}
\end{definition}\label{defUBB}

\subsection{Problem Statement}
Consider closed-loop system \eqref{CLSP}, our aim is to calculate for $ x_k $ at instant $ \tau_k $, the next optimal sampling horizon $ \sigma_k $ which will be applied to the sampling mechanism in order to ensure global uniform ultimate boundedness of the system's solution.

\subsection{Proposed self triggering mechanism and Stability Analysis: Online case}\label{STM-off-p}

\subsubsection{Self Triggering Mechanism and Stability Analysis}
Proposition \ref{STMS-off-p} ensures global uniform ultimate boundedness of the discretized system \eqref{CLSP} in which the next sampling horizon $ \sigma_k $ defined by proposed self-triggering mechanism.

In the rest of the paper, we use the notation $ \mathcal{E}(P,\upsilon)$ to refer to the ellipsoid
\begin{equation}
	\mathcal{E}(P,\upsilon)=\bigg\{x \in \mathbb{R}^n: x^T Px \leq  \upsilon	\bigg\}.
\end{equation}

For a positive definite matrix $ P \in \mathbb{R}^{n \times n} $ and a positive scalar $ \upsilon $. For all positive scalar $ r $, we denote by $ \mathcal{B}(0,r) $ the ball of radius $ \sqrt{r} $:
\begin{equation}\label{ball}
	\mathcal{B}(0,r)=\bigg\{x \in \mathbb{R}^n: 
	\Big\| x \Big\|^2_2 \leq r	\bigg\}.
\end{equation}
\begin{proposition}\label{STMS-off-p}
	Consider scalars $ \beta>0$ and $\gamma>0 $. If there exists $ \sigma^* \in S^{l_{max}}_{l_{min}}(\Gamma) $ and  symmetric positive definite matrices $ P,~M$ such that 
	
	\begin{equation} \label{LMI1-STM-on-p}
		\begin{split}	
			&\Phi_{\sigma^*}^T(P+M)\Phi_{\sigma^*} +\Big(e^{(-\beta\sum\nolimits_{j=1}^{|{\sigma^*} |}T_{\sigma^*}^{j }) }-\gamma\Big)P \succeq 0,
		\end{split} 
	\end{equation}
	and
	\begin{equation} \label{LMI2-STM-on-p}
		\begin{bmatrix} M &  P\\ P &  \frac{\gamma}{\chi}I- P \end{bmatrix} \succeq 0, 
	\end{equation}
	are verified with $ \chi =\Big(\varpi  \sum_{q=0}^{|\sigma^*|-1}C^q \Big)^2 $, then the system \eqref{CLSP} is GUUB with the self-triggering mechanism defined by
	
	\begin{equation}\label{STM-on-p}
		\begin{split}
			\sigma_k \in \left\{
			\begin{array}{ll}
				\displaystyle\argmax_{{\sigma} \in {{\bar{S}}^{l_{max}}_{l_{min}}}(\Gamma,x_k)} \Bigg\{ \frac{\sum\nolimits_{j=1}^{|{\sigma} |}T_{\sigma}^{j }}{|{\sigma} |} \Bigg\}, & If~x_k \not\in \mathcal{E}(P,1), \\ \\
				\Big\{\Big(T_{max}\Big)\Big\}, &If~ x_k \in \mathcal{E}(P,1),
			\end{array}
			\right.
		\end{split}
	\end{equation}
	with
	
	\begin{equation}\label{es-on-p}
		\begin{split}
			&{{\bar{S}}^{l_{max}}_{l_{min}}}(\Gamma,x_k) =\Bigg \{\sigma \in S^{l_{max}}_{l_{min}}(\Gamma);~\begin{pmatrix}
				x_k \\
				1
			\end{pmatrix}^T U_{\sigma} \begin{pmatrix}
				x_k \\
				1
			\end{pmatrix} \geq 0 \Bigg \},
		\end{split}
	\end{equation}
	\begin{equation} \label{U_sigma}
		\begin{split}
			U_{\sigma}=\text{diag}\Bigg(-\Phi^T_{{\sigma}} \Big(P+M\Big)\Phi_{{\sigma}}+\Big(e^{(-\beta\sum\nolimits_{j=1}^{|{\sigma} |}T_{\sigma}^{j }) }-\gamma\Big) P \\
			,-\Big(\varpi  \sum_{q=0}^{|\sigma|-1}C^q \Big)^2\lambda_{max}(P M^{-1}P+P)+\gamma  \Bigg),
		\end{split}
	\end{equation}
	and
	\begin{equation}
		T_{max}=\max \{T\in\Gamma\}.
	\end{equation}			
	Furthermore, the solution of \eqref{CLSP} converges and remains in $ \mathcal{E}(P,\mu) $, with 
	\begin{equation}
		\mu= \lambda_{max}(P) \Big(\frac{C'}{\lambda_{min}(P)}+ \varpi\Big)^2 ,
	\end{equation} 
	where $ C'=\Big\| \tilde{A}_{(T_{max})} \Big\|_2 $  and $ \varpi $ is given in Assumption \ref{pertbounded}.	  
\end{proposition}

\begin{proof}
	See Appendix.
\end{proof}

\begin{remark}
	To obtain the smallest ellipsoid $ \mathcal{E}(P,\mu) $ by which the solution of \eqref{CLSP} is bounded, one can consider a ball $ \mathcal{B}(0,\psi) $ tangent to the ellipsoid $ \mathcal{E}(P,\mu) $ (See Fig.~{ \ref{MatP2}}) such that if there exists $ \sigma^* \in S^{l_{max}}_{l_{min}}(\Gamma) $ and  symmetric positive definite matrices $ P,~M$ such that 
	\begin{equation}\label{LMI1_Per_On} 
		\begin{split}	
			&\Phi_{\sigma^*}^T(P+M)\Phi_{\sigma^*} +\Big(e^{(-\beta\sum\nolimits_{j=1}^{|{\sigma^*} |}T_{\sigma^*}^{j }) }-\gamma\Big)P \succeq 0,
		\end{split} 
	\end{equation}
	and
	\begin{equation}\label{LMI2_Per_On} 
		\begin{bmatrix} M &  P\\ P &  \frac{\gamma}{\chi}I- P \end{bmatrix} \succeq 0, 
	\end{equation}
	are verified and for the horizon $ \sigma^* $:
	
	\begin{equation}
		\begin{split}
			&\min_{\psi \in \mathbb{R}\setminus \{0\}} \psi \\
			& \text{s.t.}~ P \succ \underline{\zeta}I,~\underline{\zeta} > \frac{\mu}{\psi}.
		\end{split}
	\end{equation}
	\begin{figure}[!h] 
		\centering
		\subfloat{%
			\resizebox*{5cm}{!}{\includegraphics{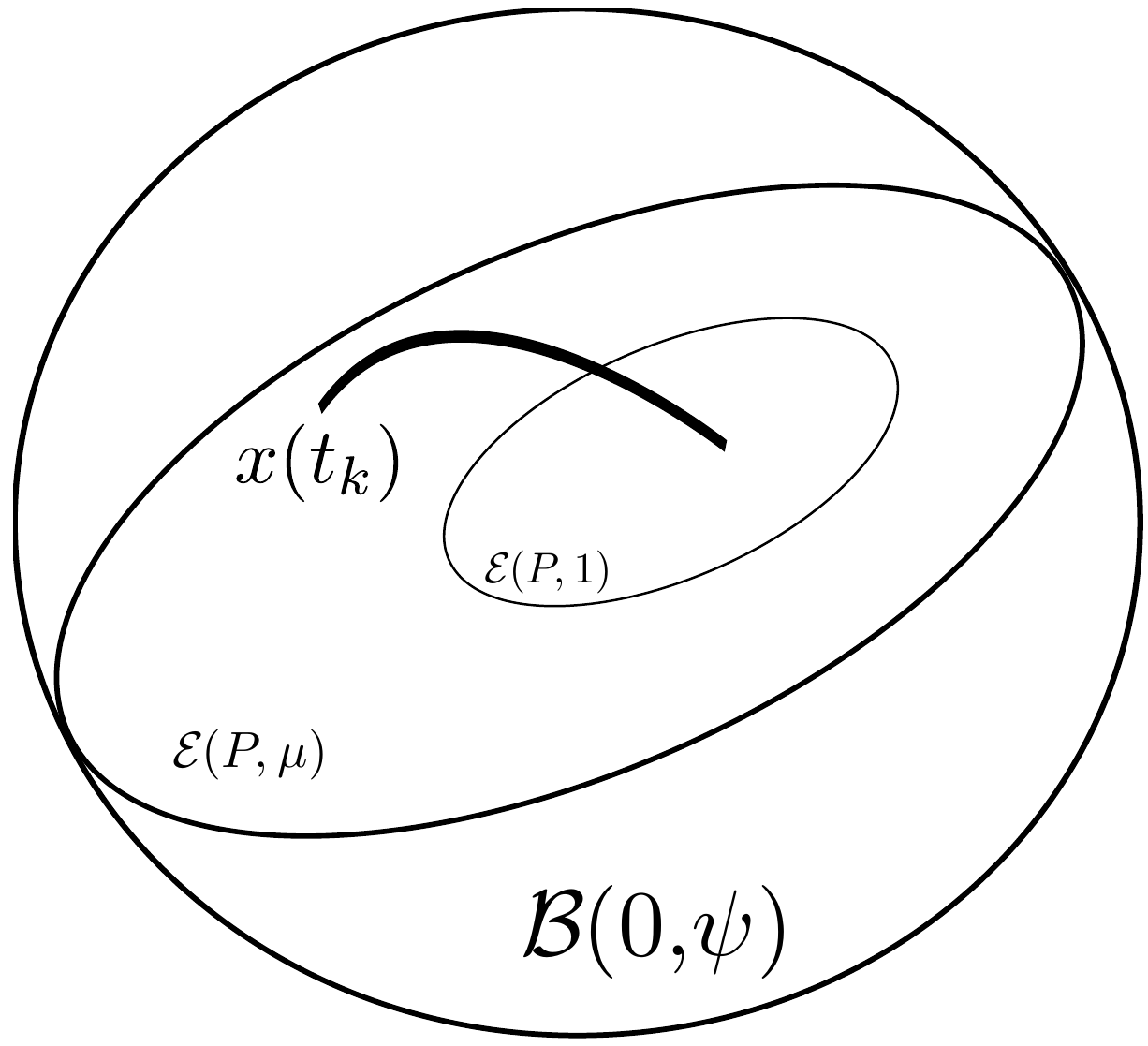}}}\hspace{2pt}
		\caption{ The illustration of three regions $ \mathcal{E}(P,1) $, $ \mathcal{E}(P,\mu) $, and $ \mathcal{B}(0,\psi) $.} 	\label{MatP2}
	\end{figure}
\end{remark}

\subsubsection{Implementation Algorithm}
Algorithm \ref{Alg3} provides a self triggering algorithm for the sampled-date system \eqref{CLSP} which generates a next optimal sampling horizon $ {\sigma}_k $ based on $ x(\tau_k) $ such that the selected sampling horizon ensures the system's solution is GUUB.

\begin{algorithm}[!htbp]
	\caption{Self Triggering Mechanism for Sampled-Data System \eqref{CLSP}: Online case}\label{Alg3}
	\begin{algorithmic}
		\STATE
		\STATE \textbf{\underline{OFFLINE Procedure:}}
		\begin{itemize}
			\STATE Define $ l_{min} $ and $ l_{max} $, the minimum and maximum length of the considered sampling horizons.
			\STATE Select positive constant $ \gamma $.
			\STATE Define a finite set $ \Gamma=\{T^1,\cdots,T^m\} $ of sampling intervals.
			\STATE Design the set $ S^{l_{max}}_{l_{min}}(\Gamma) $ of sampling horizons of length $l \in [l_{min},l_{max}] $ constituted of sampling intervals in $ \Gamma $.
			\STATE Compute the matrices $ P=P^T \succ 0 $ and \\ $ M=M^T\succ0 $ for a given stable sequence $ \sigma^*\in S^{l_{max}}_{l_{min}}(\Gamma) $ i.e., $ \Phi_{\sigma^*} $ is a Schur matrix satisfying: $\Phi_{\sigma^*}^T(P+M)\Phi_{\sigma^*} +\left(e^{(-\beta\sum_{j=1}^{|{\sigma^*} |}T_{\sigma^*}^{j }) }-\gamma\right)P \succeq 0$, and $\begin{bmatrix} M & P\\ P & \frac{\gamma}{\chi} I - P \end{bmatrix} \succeq 0$ where $ \chi =\left(\varpi \sum_{q=0}^{|\sigma^*|-1}C^q \right)^2 $.
		\end{itemize}
		\STATE
		\STATE \textbf{\underline{ONLINE Procedure:} At each instant $ \tau_k $:}
		\STATE $ {T_{avg}}_{max}=\frac{\sum_{j=1}^{|{\sigma^*} |}T_{\sigma^*}^{j }}{|{\sigma^*} |} $
		\STATE $ \sigma_{opt}=\{\sigma^*\} $
		\FORALL{$ \sigma \subset S^{l_{max}}_{l_{min}}(\Gamma) $}
		\IF{$ \begin{pmatrix} x_k \\ 1 \end{pmatrix}^T U_{\sigma} \begin{pmatrix} x_k \\ 1 \end{pmatrix}\geq 0 $}
		\STATE $ T_{avg}=\frac{\sum_{j=1}^{|{\sigma} |}T_{\sigma}^{j}}{|{\sigma} |}$ 
		\IF {$ T_{avg} > {T_{avg}}_{max}$}
		\STATE $ {T_{avg}}_{max}= T_{avg}$
		\STATE $ {\sigma_{opt}}=\{\sigma\} $
		\ELSIF {$ T_{avg} ={T_{avg}}_{max}$}
		\STATE $ {\sigma_{opt}}=\{\sigma_{opt},\sigma\} $
		\ENDIF
		\ENDIF
		\ENDFOR
		\STATE Choose randomly  $ \sigma_k $, from  $ {\sigma_{opt}} $.
	\end{algorithmic}
\end{algorithm}

\subsection{Proposed self triggering mechanism and Stability Analysis: Offline case}
Similar to the Section \ref{wp-online}, an offline tractable method is used for splitting of the state space into regions of user-defined number to compute an optimal possible stablizing horizon based on a region that the state of the system \eqref{CLSP} is belong to such that ensures globally uniformly ultimately  boundedness of the solution.
\subsubsection{Self Triggering Mechanism and Stability Analysis} 
Proposition \ref{STMS-on-p} ensures globally uniformly ultimately  boundedness of the discretized system \eqref{CLSP} in which the next optimal sampling horizon $ \Psi_{(c)} $ defined by proposed self-triggering for a conic region $ \mathcal{R}_c, c \in \mathbb{N} $.
\begin{proposition}\label{STMS-on-p}
	Consider scalars $ \beta>0$, $\gamma_1>0 $, and $\gamma_2>0 $. If there exists $ \sigma^* \in S^{l_{max}}_{l_{min}}(\Gamma) $ and  symmetric positive definite matrices $ P$ such that 
	\begin{equation} \label{SigmaOmega-off}
		\begin{split}
			& U=\\
			&\begin{pmatrix}
				-\Phi_{\sigma^*}^TP\Phi_{\sigma^*}+\Big(\bar{\beta}-\gamma_1\Big)P & * & *\\
				-P\Phi_{\sigma^*} & \frac{\gamma_2}{\chi}I-P & *\\
				\textbf{0} & \textbf{0} &  {-\gamma_2+\gamma_1} 
			\end{pmatrix}\succeq 0 ,			
		\end{split} 
	\end{equation}
	in which $ \chi =\varpi \sum_{q=0}^{|\sigma^*|-1}C^q $ and $ \bar{\beta}=e^{(-\beta\sum\nolimits_{j=1}^{|{\sigma^*} |}T_{\sigma^*}^{j }) } $, then the system \eqref{CLSP} is GUUB with the self-triggering mechanism defined by
	\begin{equation}\label{STMoff-p}
		\begin{split}
			\sigma_k \in \left\{
			\begin{array}{l}
				 If~x_k \not\in \mathcal{E}(P,1):\\
				\displaystyle\argmax_{\sigma \in \Psi_{(c)}:~ c \in \{1,\cdots,N\}~\text{and}~x_k \in \mathcal{R}_c} \Bigg\{ \frac{\sum\nolimits_{j=1}^{|{\sigma} |}T_{\sigma}^{j }}{|{\sigma} |}  \Bigg\},  \\ \\
				If~x_k \in \mathcal{E}(P,1):\\
				\Big\{\Big(T_{max}\Big)\Big\},
			\end{array}
			\right.
		\end{split}
	\end{equation} 
	with
	\begin{equation} 
		\begin{split}
			& \Psi_{(c)}  = \argmax_{{\sigma} \in {{\bar{S}}^{l_{max}}_{l_{min}}}(\Gamma,\mathcal{R}_c)}\Bigg\{ \frac{\sum\nolimits_{j=1}^{|{\sigma} |}T_{\sigma}^{j }}{|{\sigma} |}  \Bigg\},
		\end{split}
	\end{equation}
	where
	\begin{equation}\label{es-off-p}
		\begin{split}
			&{{\bar{S}}^{l_{max}}_{l_{min}}}(\Gamma,\mathcal{R}_c) =\Big\{\sigma \in S^{l_{max}}_{l_{min}}(\Gamma),~\epsilon_c>0;~U_{c}\succeq 0 \Big\},
		\end{split}
	\end{equation}
	and $ U_{c}=[u_{ij}] $ is a symmetric matrix defined by
	\begin{equation}
		\begin{split}		
			&u_{11}=\epsilon_c Q_c-\Phi_{\sigma}^T P \Phi_{\sigma}+\bar{\beta} P-\gamma_1 P,\\ &u_{22}=\frac{\gamma_2}{\varpi \sum_{q=0}^{|\sigma|-1}C^q}I-P,\\ &u_{33}={-\gamma_2+\gamma_1},\\
			&u_{21}=-P\Phi_{\sigma},\\
			&u_{31}=\textbf{0},\\
			&u_{32}=\textbf{0} .
		\end{split}
	\end{equation}
	and
	\begin{equation}
		T_{max}=\max \{T\in\Gamma\}.
	\end{equation}			
	Furthermore, the solution of \eqref{CLSP} converges and remains in $ \mathcal{E}(P,\mu) $, with 
	\begin{equation}
		\mu= \lambda_{max}(P) \Big(\frac{C'}{\lambda_{min}(P)}+ \varpi\Big)^2 ,
	\end{equation} 
	where $ C'=\Big\| \tilde{A}_{(T_{max})} \Big\|_2 $  and $ \varpi $ is given in Assumption \ref{pertbounded}.	
	
\end{proposition}
\begin{proof}
	See Appendix.
\end{proof}
\subsubsection{Implementation Algorithm}
Algorithm \ref{Alg4} provides a self triggering mechanism for the sampled-date system \eqref{CLSP} which generates the next stabilizing optimal sampling horizon $ {\sigma} \in \Psi_{(c)} $ based on $ x(\tau_k) \in \mathcal{R}_c $ such that the selected sampling horizon ensures globally uniformly ultimately boundedness of the solution.
\begin{algorithm}[]
	\caption{Self Triggering Mechanism for Sampled Data System \eqref{CLSP}: Offline case}\label{Alg4}
	\begin{algorithmic} 
		\STATE
		\STATE \textbf{\underline{OFFLINE Procedure:}}
		\begin{itemize}
			\STATE Define $N$: number of conic regions.
			\STATE Define $l_{max}$ and $l_{min}$, the minimum and maximum length of the considered sampling horizons.
			\STATE Select positive constants $\gamma_1$ and $\gamma_2$.
			\STATE Define a finite set $\Gamma=\{T^1,\cdots,T^m\}$ of sampling intervals.
			\STATE Design the set $S^{l_{max}}_{l_{min}}(\Gamma)$ of sampling horizons of length $l \in [l_{min},l_{max}]$ constituted of sampling intervals in $\Gamma$.
			\STATE Compute the matrix $P=P^T\succ0$ for a given stable sequence $\sigma^*\in S^{l_{max}}_{l_{min}}(\Gamma)$ i.e., $\Phi_{\sigma^*}$ is a Schur matrix satisfying:\\ $\begin{pmatrix} -\Phi_{\sigma^*}^TP\Phi_{\sigma^*}+(\bar{\beta}-\gamma_1)P & * & *\\ -P\Phi_{\sigma^*} & \frac{\gamma_2}{\chi}I_2-P & *\\ \textbf{0} & \textbf{0} & {-\gamma_2+\gamma_1} \end{pmatrix}\succeq 0$, in which $\chi =\varpi \sum_{q=0}^{|\sigma^*|-1}C^q$ and $\bar{\beta}=e^{(-\beta\sum\nolimits_{j=1}^{|{\sigma^*}|}T_{\sigma^*}^{j})}$.
			\STATE Design a matrix $Q_c$ for each conic region $\mathcal{R}_c=\{x \in \mathbb{R}^n: x^T Q_c x \geq 0\},~\forall c \in \{1,\cdots,N\}$.
			\STATE Compute the set $\Psi_{(c)}$ of optimal sampling horizons for each conic region $\mathcal{R}_c$:
		\begin{itemize}

		\FORALL{$c=1:N$}
		\STATE ${T_{avg}}_{max}=\frac{\sum\nolimits_{j=1}^{|{\sigma^*}|}T_{\sigma^*}^{j}}{|{\sigma^*}|}$
		\STATE ${\Psi^{opt}_{(c)}}=\{\sigma^*\}$
		\FORALL{$\sigma \subset S^{l_{max}}_{l_{min}}(\Gamma)$:}
		\IF{$U \succeq 0$}
		\STATE $T_{avg}=\frac{\sum\nolimits_{j=1}^{|{\sigma}|}T_{\sigma}^{j}}{|{\sigma}|}$
		\IF{$T_{avg} > {T_{avg}}_{max}$}
		\STATE ${T_{avg}}_{max}= T_{avg}$
		\STATE ${\Psi^{opt}_{(c)}}=\{\sigma\}$
		\ELSIF{$T_{avg} ={T_{avg}}_{max}$}
		\STATE ${\Psi^{opt}_{(c)}}=\{{\Psi^{opt}_{(c)}},\sigma\}$
		\ENDIF
		\ENDIF
		\ENDFOR
		\STATE $\Psi_{(c)}={\Psi^{opt}_{(c)}}$
		\ENDFOR

		\end{itemize}
		\end{itemize}
		\STATE \textbf{\underline{ONLINE Procedure:} At each instant $\tau_k$:}
		\begin{itemize}
			\STATE Determine the conic region $\mathcal{R}_c,~c \in \{1,\cdots,N\}$, in which current state $x_k$ belongs.
			\STATE Select randomly a corresponding optimal horizon $\sigma_k$ from the sets $\Psi_{(c)},~c \in \{1,\cdots,N\}$, for which $x_k \in \mathcal{R}_c$.
		\end{itemize}
	\end{algorithmic}
\end{algorithm}

\section{Simulation}\label{Simulation}
\subsection{Perturbation-free Case}
\subsubsection{Proposed Self Triggering Mechanism: Online Procedure}
To verify the proposed approach, we consider a continuous time LTI control system \eqref{syst} and a sampled-date sate-feedback controller \eqref{statefc} with 
\begin{equation} \label{syssim}
	A=\begin{bmatrix}
		0&1\\-2&3
	\end{bmatrix}, ~ 
	B=\begin{bmatrix}
		0\\1
	\end{bmatrix}, ~ 
	K=\begin{bmatrix}
		1&-4
	\end{bmatrix}
\end{equation}

Consider sampling horizon of minimum length $ l_{min}=1 $ and maximum length $ l_{max}=6 $, a set of sampling intervals $ \Gamma=\{\begin{matrix}
	0.5 & 0.75& 1& 1.25 &1.5 &1.75& 2 &2.25& 2.75
\end{matrix}\} $, and the initial state $ x_0=[\begin{matrix} 5& -2 \end{matrix}]$.

\begin{figure}[h] 
	\centering
	\subfloat[Evolution of the system's states]{%
		\resizebox*{4.3cm}{!}{\includegraphics{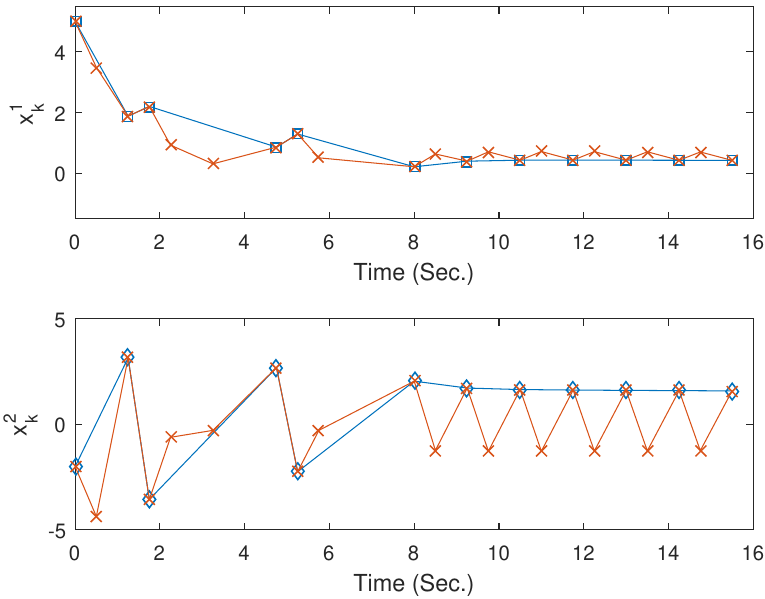}}}
	\subfloat[Evolution of the Lyapunov function]{%
		\resizebox*{4.3cm}{!}{\includegraphics{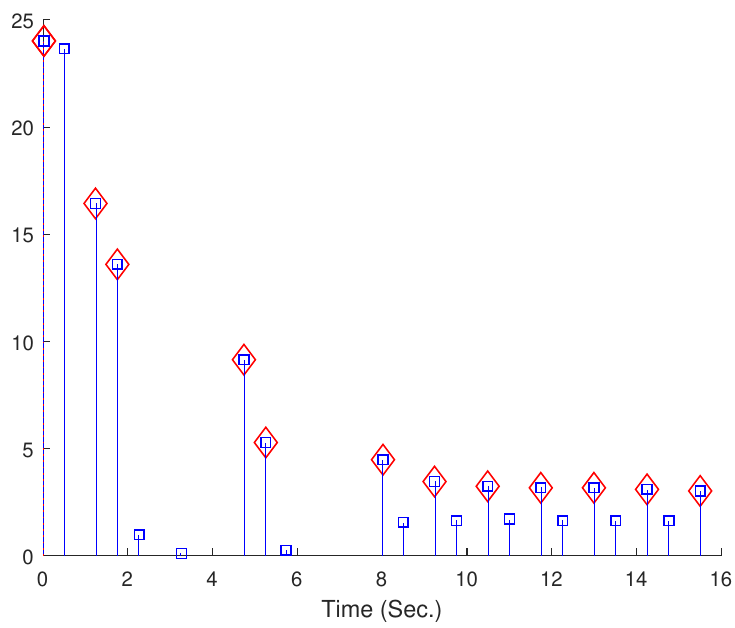}}}
	\subfloat[Steps variation and sampling horizon sequences]{%
		\resizebox*{4.3cm}{!}{\includegraphics{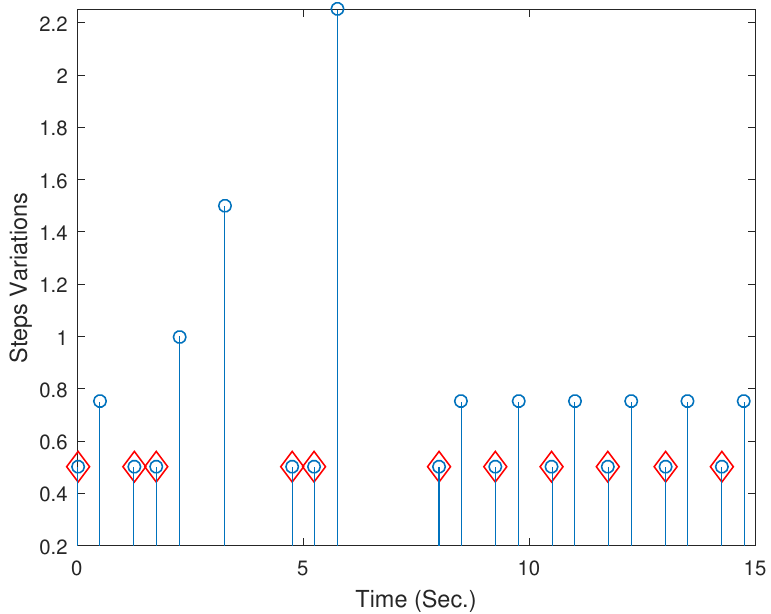}}}
	\caption{ Step Variations, Evolution of the system's states and Lyapunov function using variable sampling steps and horizons with a decay rate $ \beta=0 $} 	\label{SS1}
\end{figure}

\begin{figure}[h] 
	\centering
	\subfloat[Evolution of the system's states]{%
		\resizebox*{4.3cm}{!}{\includegraphics{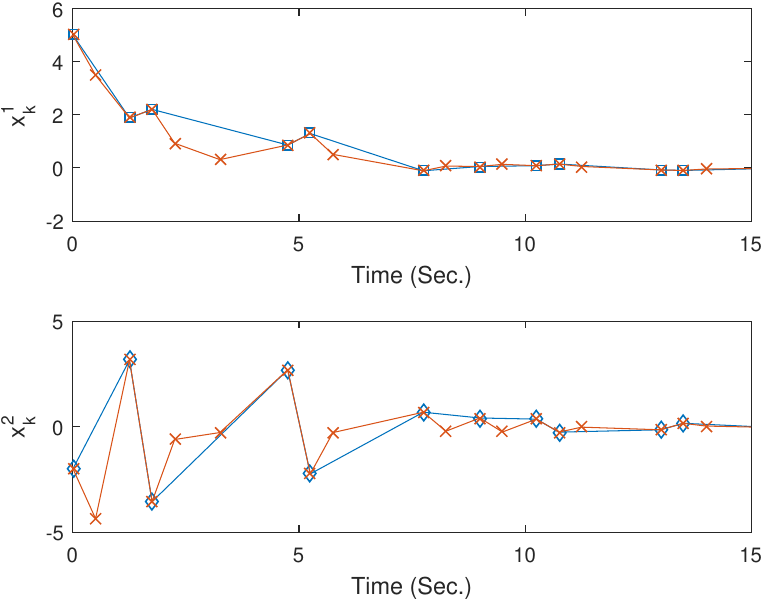}}}
	\subfloat[Evolution of the Lyapunov function]{%
		\resizebox*{4.3cm}{!}{\includegraphics{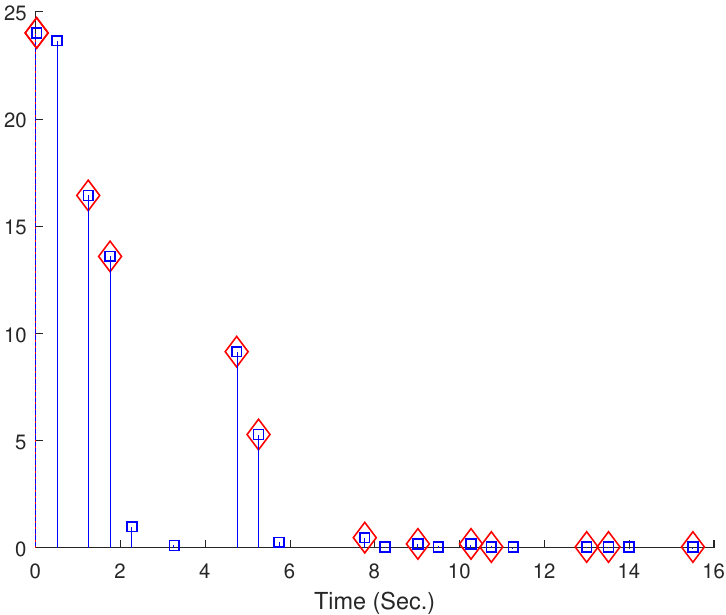}}}
	\subfloat[Steps variation and sampling horizon sequences]{%
		\resizebox*{4.3cm}{!}{\includegraphics{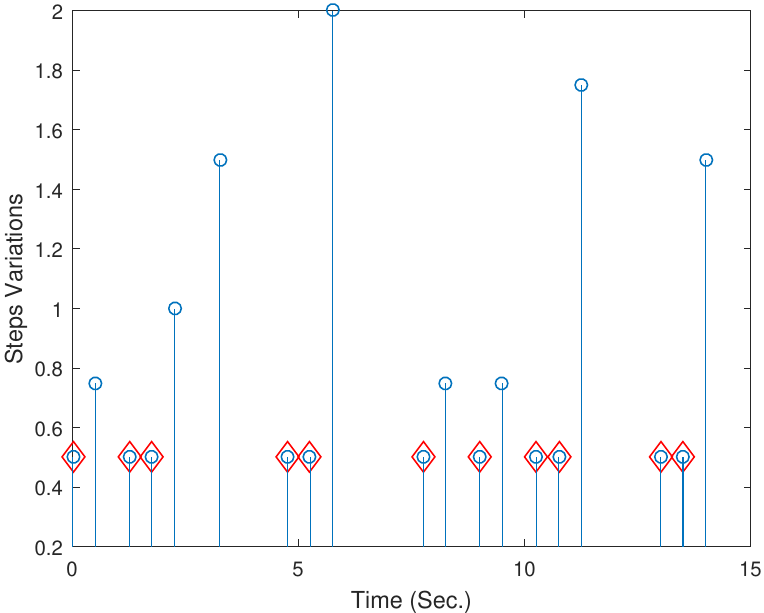}}}
	\caption{ Step Variations, Evolution of the system's states and Lyapunov function using variable sampling steps and horizons with a decay rate $ \beta=0.1 $} 	\label{SS2}
\end{figure}

Step Variations, evolution of the system’s states and Lyapunov function using variable sampling intervals with a decay rate $ \beta=0 $ (resp. $ \beta=0.1 $) are shown in Fig.~{ \ref{SS1}} (resp. Fig.~{ \ref{SS2}}). We observe that the average of sampling intervals is $ 1.4091 $ (resp. $ 1.4091 $ ) which is considerably greater than $ T_{max} $. In addition, we see that non-stabilizing sampling steps are used to ensure the asymptotic/exponential stability of the system.

\subsubsection{Proposed Self Triggering Mechanism: Offline Procedure}
\begin{figure}[h] 
	\centering
	\subfloat[Evolution of the system's states]{%
		\resizebox*{4.3cm}{!}{\includegraphics{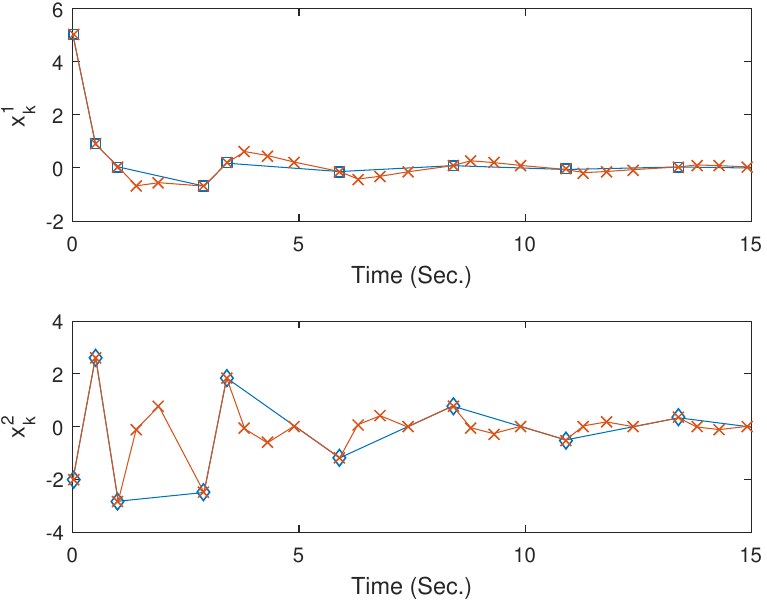}}}
	\subfloat[Evolution of the Lyapunov function]{%
		\resizebox*{4.3cm}{!}{\includegraphics{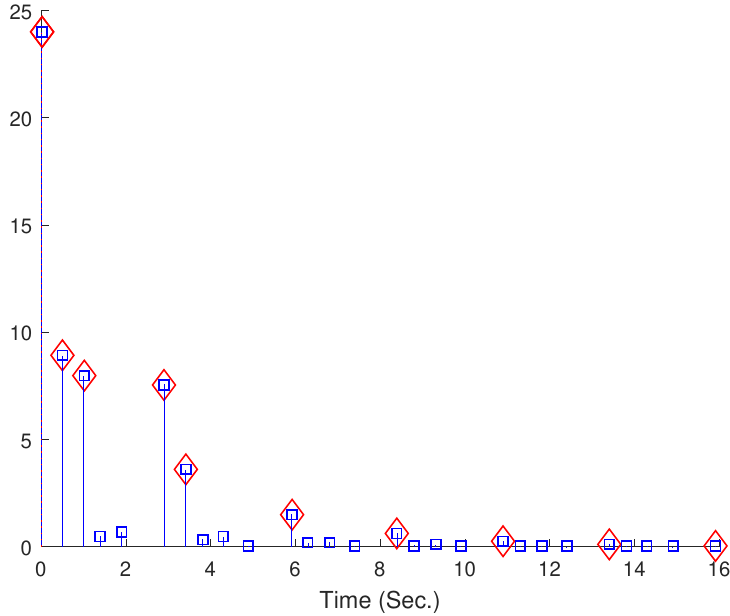}}}
	\subfloat[Steps variation and sampling horizon sequences]{%
		\resizebox*{4.3cm}{!}{\includegraphics{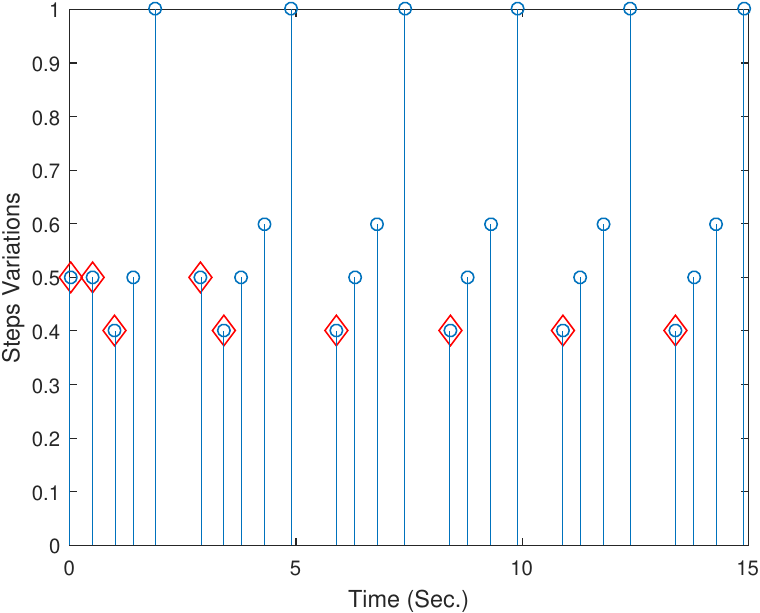}}}
	\caption{ Step Variations, Evolution of the system's states and Lyapunov function using variable sampling steps and horizons with a decay rate $ \beta=0 $} 	\label{SS3}
\end{figure}

\begin{figure}[h] 
	\centering
	\subfloat[Evolution of the system's states]{%
		\resizebox*{4.3cm}{!}{\includegraphics{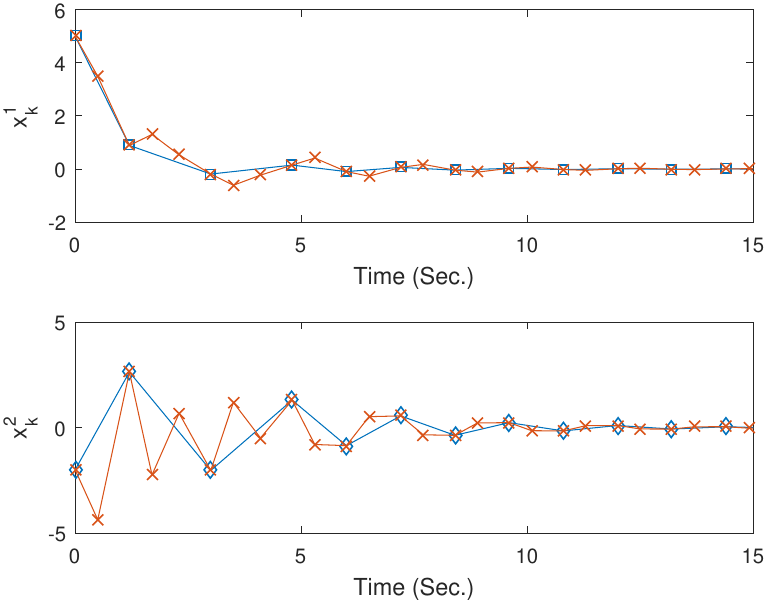}}}
	\subfloat[Evolution of the Lyapunov function]{%
		\resizebox*{4.3cm}{!}{\includegraphics{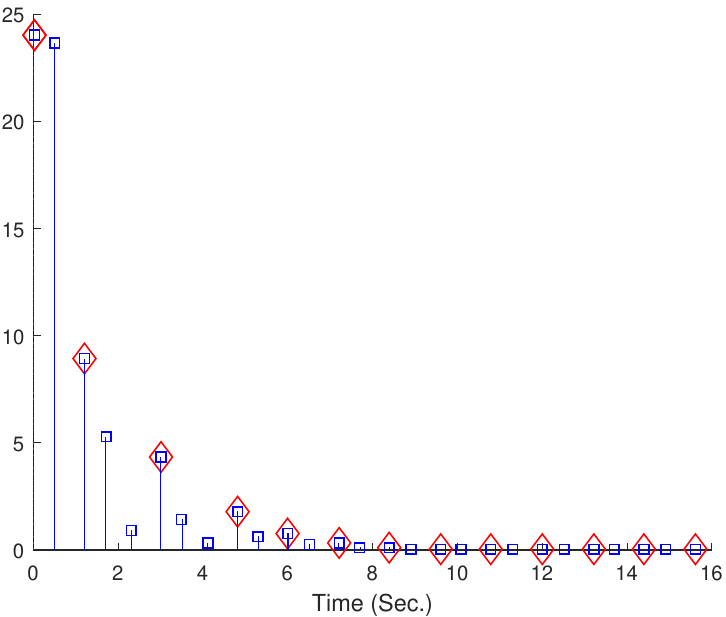}}}
	\subfloat[Steps variation and sampling horizon sequences]{%
		\resizebox*{4.3cm}{!}{\includegraphics{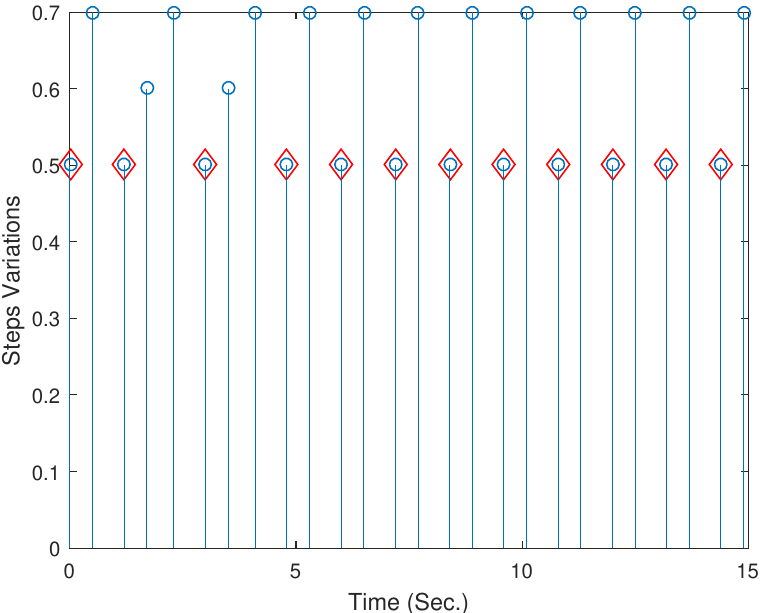}}}
	\caption{ Step Variations, Evolution of the system's states and Lyapunov function using variable sampling steps and horizons with a decay rate $ \beta=0.1 $} 	\label{SS4}
\end{figure}

After setting the number of equal conic regions $ N=100 $, minimum length $ l_{min}=1 $ and maximum length $ l_{max}=6 $, a set of sampling intervals $ \Gamma=\{\begin{matrix}
	0.2 & 0.3 & 0.4& 0.5& 0.6 &0.7
\end{matrix}\} $, and the initial state $ x_0=[\begin{matrix} 10 & -15 \end{matrix}]$, we obtain a mapping of the state-space that gives an optimal sampling horizon sequence which has a maximum possible average sampling interval while ensures the system's exponential stability for a given decay rate $ \beta>0 $ using the Algorithm \ref{Alg2}. Fig.~{ \ref{SS3}} (resp. Fig.~{ \ref{SS4}}) shows simulation results with $ \beta=0 $ (resp.$ \beta=0.1 $). We can see that the average of sampling intervals is $ 2.06 $ (resp. $ 1.2480 $).
\subsection{Perturbed Case}
\subsubsection{Proposed Self Triggering Mechanism: Online Procedure}
Consider a continuous time LTI control system described in \eqref{systP} with
\begin{equation*}
	\begin{split}
	& A=\begin{bmatrix}
		0&1\\-2&3
	\end{bmatrix}, ~ 
	B=\begin{bmatrix}
		0\\1
	\end{bmatrix}, ~ 
	K=\begin{bmatrix}
		1&-4
	\end{bmatrix}, ~ 
	D=\begin{bmatrix}
		1\\1
	\end{bmatrix}, \\
	& w(t)=sin(5\pi t)
\end{split}
\end{equation*}

Also, we choose sampling horizon of minimum length $ l_{min}=1 $ and maximum length $ l_{max}=6 $, a set of sampling intervals $ \Gamma=\{\begin{matrix}
	0.1& 0.2 & 0.3& 0.4& 0.5 &0.6 &0.7
\end{matrix}\} $, and the initial state $ x_0=[\begin{matrix} 5& -2 \end{matrix}]$.
The LMI in equations \eqref{LMI1_Per_On} and \eqref{LMI2_Per_On} with $ \gamma=0.35 $ yields, 
\begin{equation*}
	P=\begin{bmatrix} 1.1403  & -0.1484 \\ -0.1484  &  1.7694 \end{bmatrix}, ~  M=\begin{bmatrix} 9.5808  &  2.0305 \\ 2.0305  &  8.0881 \end{bmatrix}
\end{equation*}

\begin{figure}[h] 
	\centering
	\subfloat[Evolution of the system's states]{%
		\resizebox*{4.3cm}{!}{\includegraphics{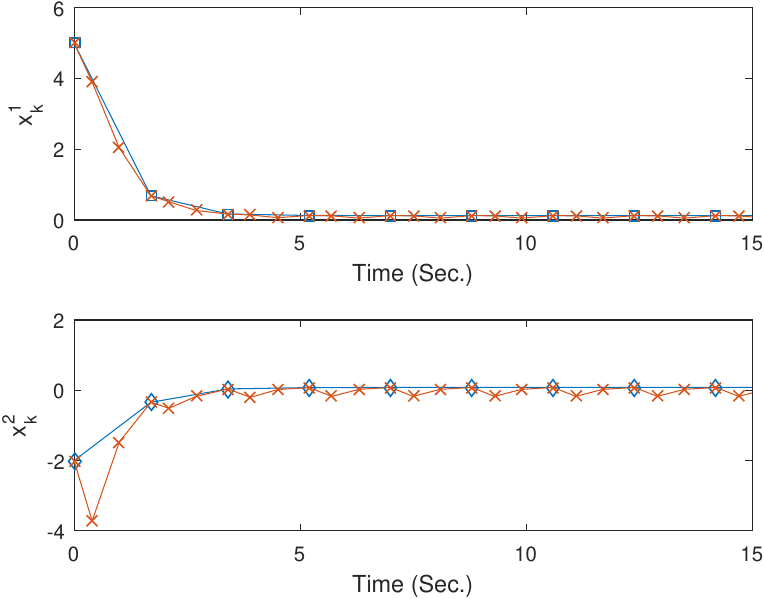}}}
	\subfloat[Evolution of the Lyapunov function]{%
		\resizebox*{4.3cm}{!}{\includegraphics{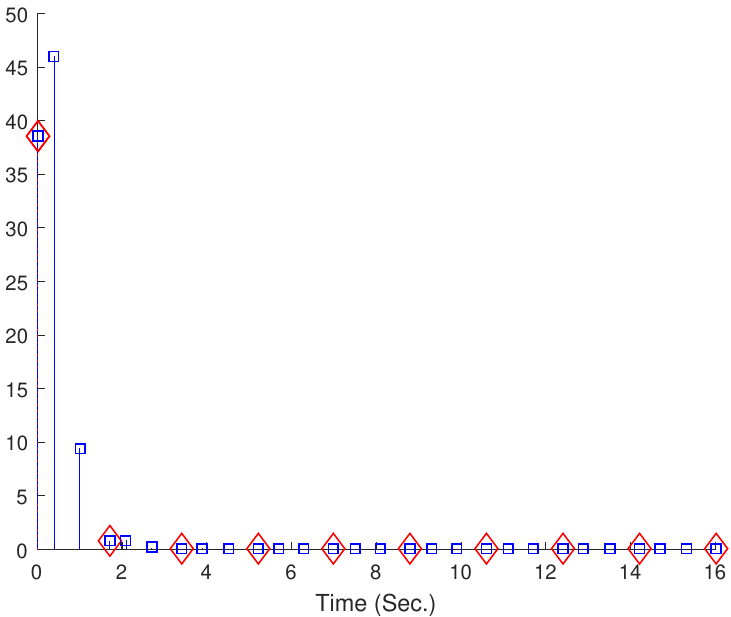}}}
	\subfloat[Steps variation and sampling horizon sequences]{%
		\resizebox*{4.3cm}{!}{\includegraphics{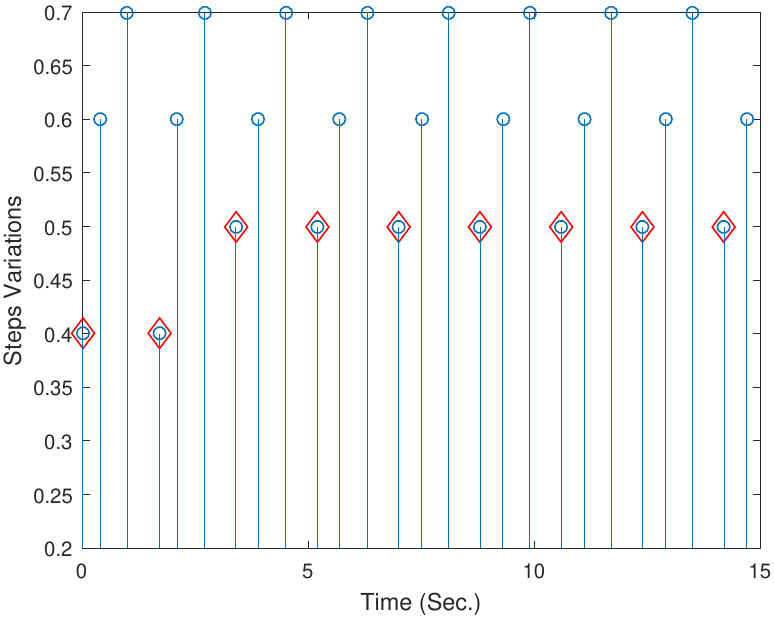}}}
	\caption{ Step Variations, Evolution of the system's states and Lyapunov function using variable sampling steps and horizons with a decay rate $ \beta=0 $} 	\label{SS5}
\end{figure}

Step Variations, evolution of the system’s states and Lyapunov function using variable sampling intervals with a decay rate $ \beta=0 $ are shown in Fig.~{ \ref{SS5}}. We can see that the average of sampling intervals is $ 1.7778 $.
To verify exponential stability of the system, consider $ \beta=0.1 $, $ \gamma=0.35 $, and  
\begin{equation*}
	P=\begin{bmatrix} 1.1529  & -0.1239 \\ -0.1239  &  1.7206 \end{bmatrix}, ~ M=\begin{bmatrix} 9.5680  &  2.0150 \\ 2.0150   & 8.0855 \end{bmatrix}
\end{equation*} 

\begin{figure}[h] 
	\centering
	\subfloat[Evolution of the system's states]{%
		\resizebox*{4.3cm}{!}{\includegraphics{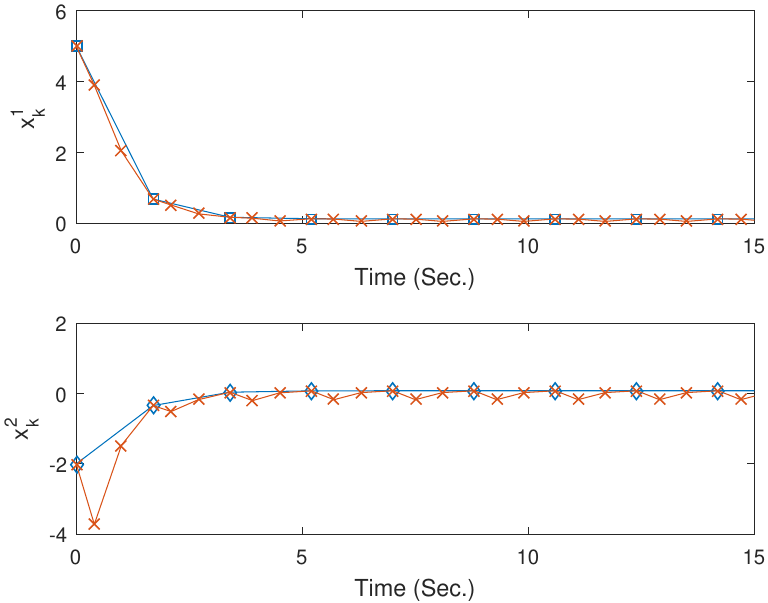}}}
	\subfloat[Evolution of the Lyapunov function]{%
		\resizebox*{4.3cm}{!}{\includegraphics{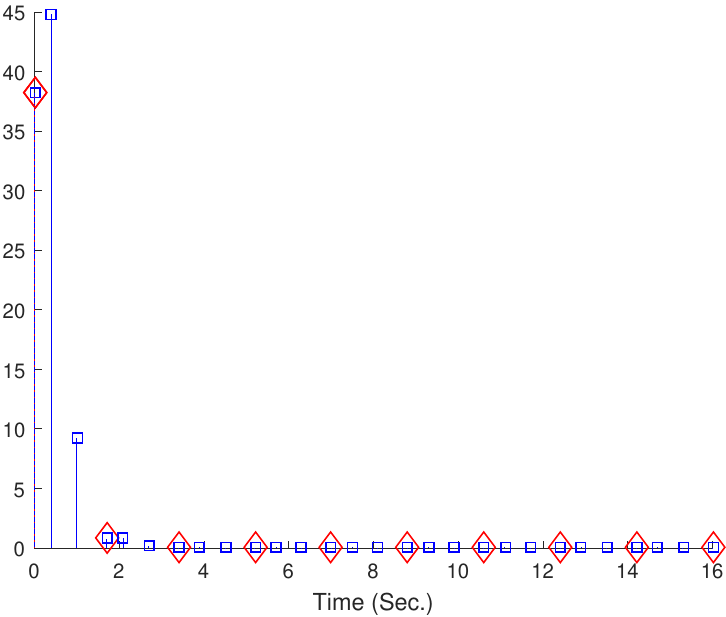}}}
	\subfloat[Steps variation and sampling horizon sequences]{%
		\resizebox*{4.3cm}{!}{\includegraphics{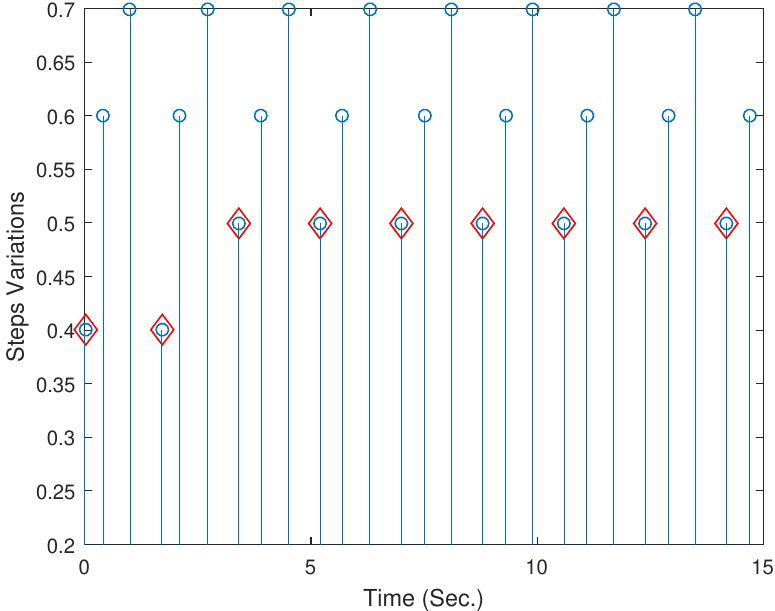}}}
	\caption{ Step Variations, Evolution of the system's states and Lyapunov function using variable sampling steps and horizons with a decay rate $ \beta=0.1 $} 	\label{SS6}
\end{figure}

Step Variations, evolution of the system’s states and Lyapunov function using variable sampling intervals are shown in Fig.~{ \ref{SS6}}. Also, the average of sampling intervals is $ 1.7778 $.

\subsubsection{Proposed Self Triggering Mechanism: Offline Procedure}
After setting the number of equal conic regions $ N=1000 $, minimum length $ l_{min}=1 $ and maximum length $ l_{max}=6 $, a set of sampling intervals $ \Gamma=\{\begin{matrix}
	0.1 & 0.2 & 0.3 & 0.4 & 0.5 & 0.6 & 0.7
\end{matrix}\} $, and the initial state $ x_0=[\begin{matrix} 2& -15 \end{matrix}]$, we obtain a mapping of the state-space.   Also, The LMI in equation \eqref{SigmaOmega-off} with $ \gamma_1=0.3 $ and $ \gamma_2=0.25 $ yields, 
\begin{equation*}
	P=\begin{bmatrix} 0.8017  &  0.5047 \\ 0.5047  &  0.4666 \end{bmatrix}
\end{equation*}

\begin{figure}[h] 
	\centering
	\subfloat[Evolution of the system's states]{%
		\resizebox*{4.3cm}{!}{\includegraphics{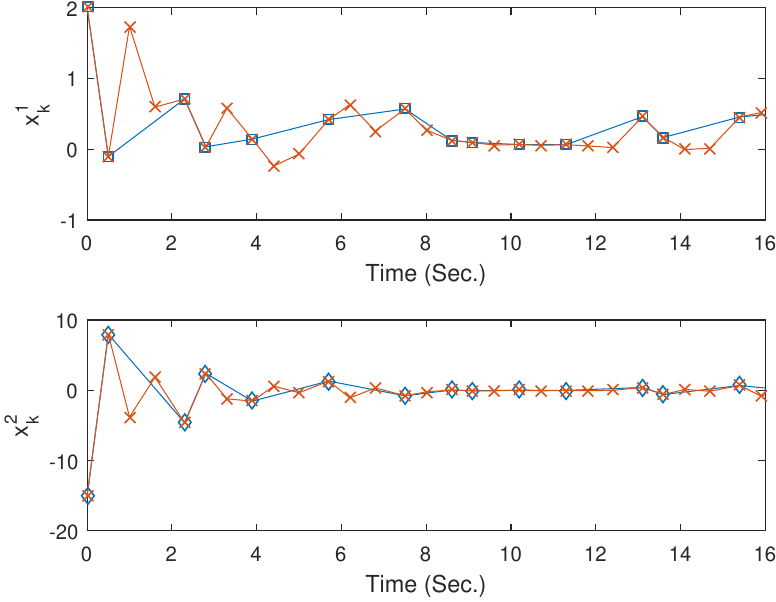}}}
	\subfloat[Evolution of the Lyapunov function]{%
		\resizebox*{4.3cm}{!}{\includegraphics{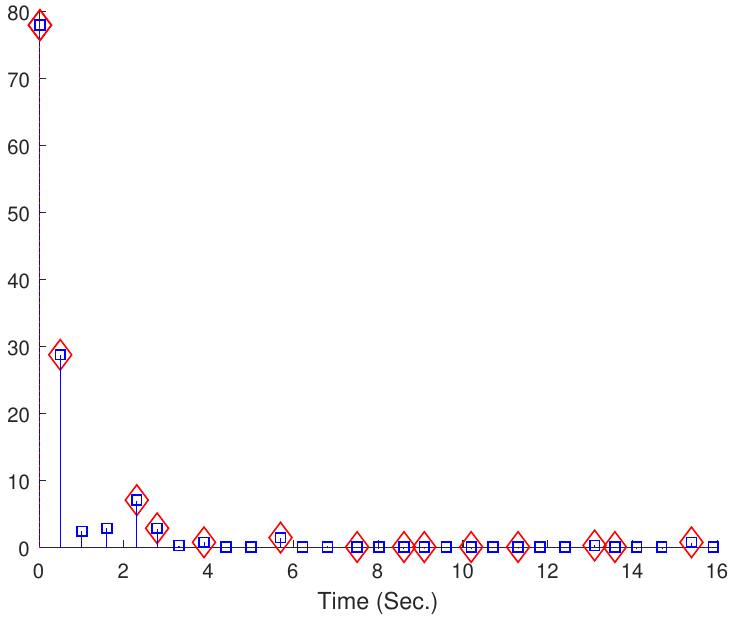}}}
	\subfloat[Steps variation and sampling horizon sequences]{%
		\resizebox*{4.3cm}{!}{\includegraphics{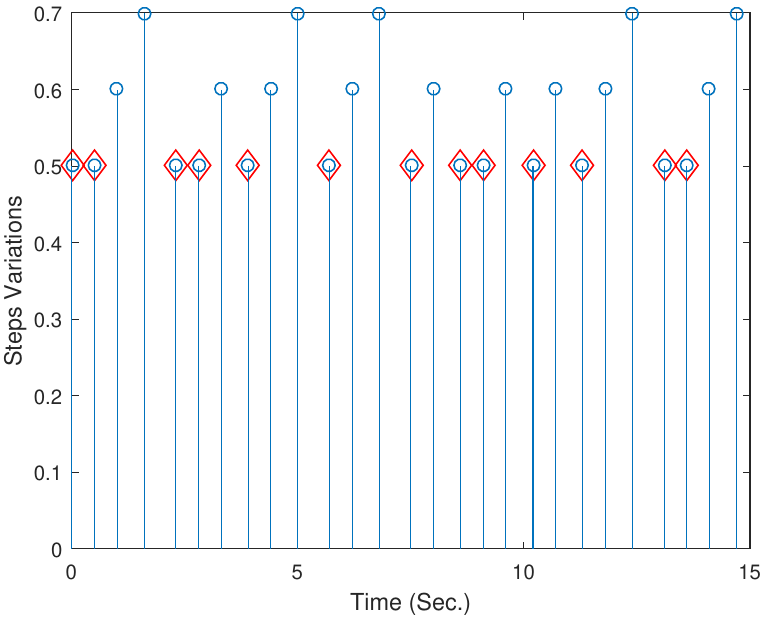}}}
	\caption{ Step Variations, Evolution of the system's states and Lyapunov function using variable sampling steps and horizons with a decay rate $ \beta=0 $.} 	\label{SS7}
\end{figure}

Fig.~{ \ref{SS7}} shows simulation results with $ \beta=0 $ using Algorithm \ref{Alg4}. We can see that the average of sampling intervals is $ 1.3087 $.
To verify exponential stability of the system, consider $ \beta=0.1 $, $ \gamma_1=0.3 $, $ \gamma_2=0.25 $, and 
\begin{equation*}
	P=\begin{bmatrix} 0.7769  &  0.4965 \\ 0.4965  &  0.4521 \end{bmatrix}
\end{equation*}

\begin{figure}[h] 
	\centering
	\subfloat[Evolution of the system's states]{%
		\resizebox*{4.3cm}{!}{\includegraphics{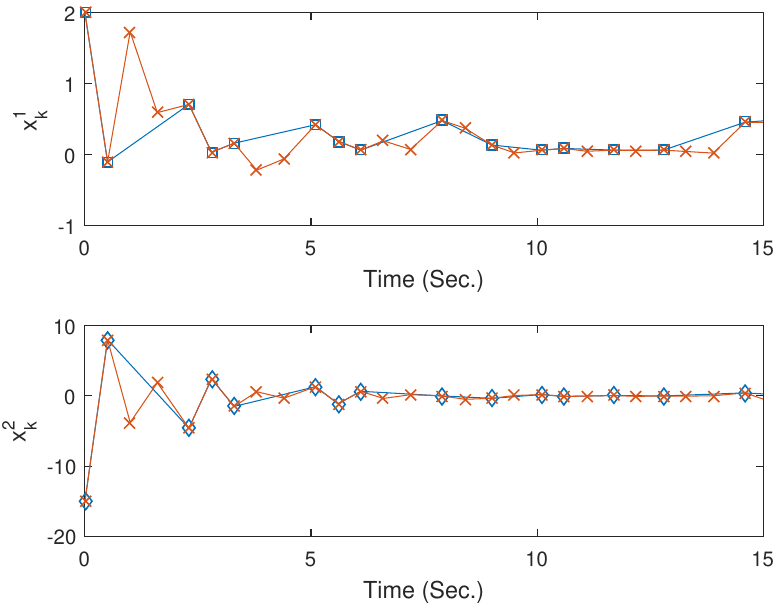}}}
	\subfloat[Evolution of the Lyapunov function]{%
		\resizebox*{4.3cm}{!}{\includegraphics{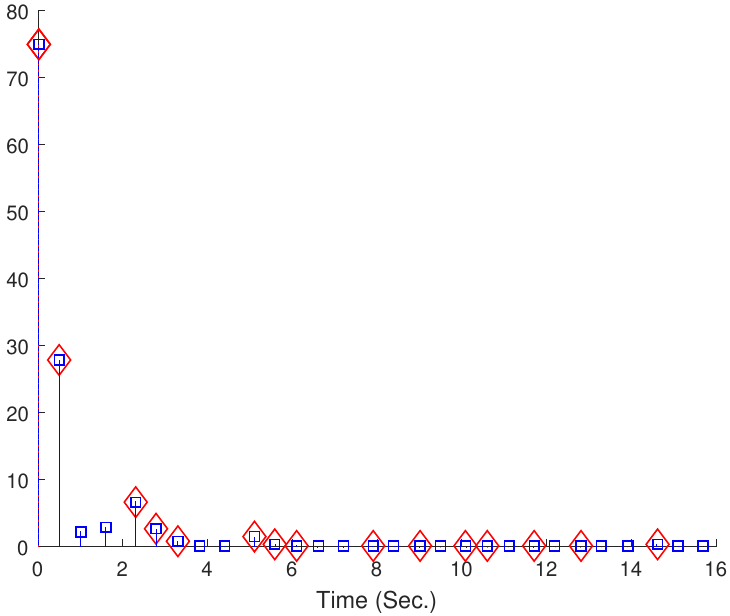}}}
	\subfloat[Steps variation and sampling horizon sequences]{%
		\resizebox*{4.3cm}{!}{\includegraphics{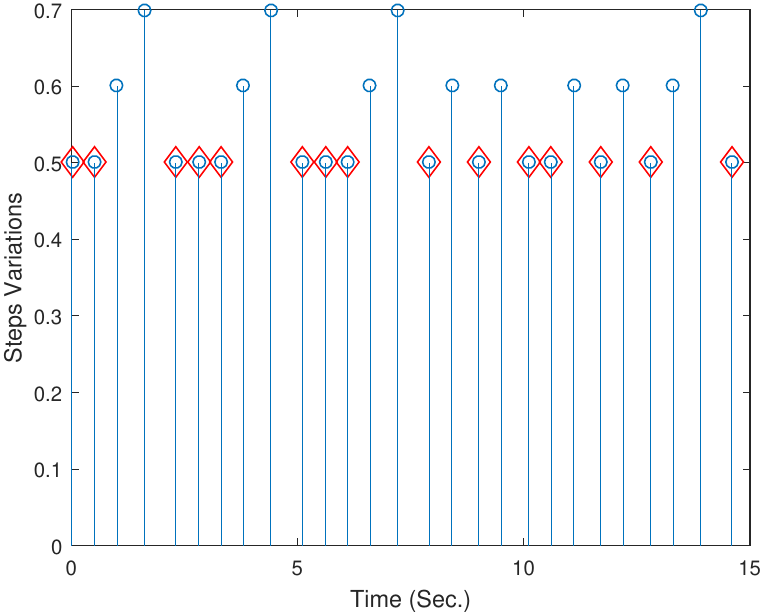}}}
	\caption{ Step Variations, Evolution of the system's states and Lyapunov function using variable sampling steps and horizons with a decay rate $ \beta=0.1 $} 	\label{SS8}
\end{figure}

Step Variations, evolution of the system’s states and Lyapunov function using variable sampling intervals are shown in Fig.~{ \ref{SS8}}. Also, the average of sampling intervals is $ 1.0655 $. 

\section{Conclusion}\label{Conclusion}

This paper introduces an optimal self-triggering strategy for linear sampled-data systems aimed at extending the sampling interval while ensuring system stability. This approach provides theoretical stability guarantees for both unperturbed and perturbed systems. We detail both the computationally intensive online version, suitable for detailed analysis and theoretical exploration, and the practical offline version, designed for real-world application and ease of implementation. The distinction emphasizes our method's adaptability to varying computational constraints and operational environments. Simulation results affirm that the proposed strategy effectively reduces the required sampling frequency while maintaining system performance, underscoring its utility in resource-constrained settings.

While the current study focuses on enhancing resource efficiency in embedded and networked control systems, future work could explore adaptive self-triggering mechanisms that respond dynamically to changing system conditions or unforeseen disturbances. This direction could further optimize resource usage and system responsiveness under variable operational scenarios. Furthermore, a self-triggered mechanism can be devised aiming at
maximizing the average of next sampling intervals for each user-defined
set of sensors individually. In other words, each user-defined set of sensors
is updated asynchronously. Finally, extending this framework to nonlinear systems would expand its applicability to an even wider range of real-world control problems.

\section*{Acknowledgments}
The author gratefully acknowledges Dr. Christophe Fiter for his significant contributions and dedicated supervision throughout this research conducted at CRIStAL in 2017--2018, and thanks Dr. Lotfi Belkoura for his valuable guidance and support.

\bibliographystyle{unsrtnat}
\bibliography{references} 

\begin{thebibliography}{39}
\providecommand{\natexlab}[1]{#1}
\providecommand{\url}[1]{\texttt{#1}}
\expandafter\ifx\csname urlstyle\endcsname\relax
  \providecommand{\doi}[1]{doi: #1}\else
  \providecommand{\doi}{doi: \begingroup \urlstyle{rm}\Url}\fi

\bibitem[Heemels et~al.(2012)Heemels, Johansson, and
  Tabuada]{heemels2012introduction}
Wilhelmus~PMH Heemels, Karl~Henrik Johansson, and Paulo Tabuada.
\newblock An introduction to event-triggered and self-triggered control.
\newblock In \emph{2012 ieee 51st ieee conference on decision and control
  (cdc)}, pages 3270--3285. IEEE, 2012.

\bibitem[Miskowicz(2018)]{miskowicz2018event}
Marek Miskowicz.
\newblock \emph{Event-based control and signal processing}.
\newblock CRC press, 2018.

\bibitem[Hetel et~al.(2017)Hetel, Fiter, Omran, Seuret, Fridman, Richard, and
  Niculescu]{hetel2017recent}
Laurentiu Hetel, Christophe Fiter, Hassan Omran, Alexandre Seuret, Emilia
  Fridman, Jean-Pierre Richard, and Silviu~Iulian Niculescu.
\newblock Recent developments on the stability of systems with aperiodic
  sampling: An overview.
\newblock \emph{Automatica}, 76:\penalty0 309--335, 2017.

\bibitem[Mikheev et~al.(1988)Mikheev, Sobolev, and
  Fridman]{mikheev1988asymptotic}
Yu~V Mikheev, VA~Sobolev, and EM~Fridman.
\newblock Asymptotic analysis of digital-control systems.
\newblock \emph{Automation and remote control}, 49\penalty0 (9):\penalty0
  1175--1180, 1988.

\bibitem[{\AA}str{\"o}m and Wittenmark(2013)]{aastrom2013adaptive}
Karl~J {\AA}str{\"o}m and Bj{\"o}rn Wittenmark.
\newblock \emph{Adaptive control}.
\newblock Courier Corporation, 2013.

\bibitem[Fridman(1992)]{fridman1992use}
E.~M. Fridman.
\newblock The use of models with aftereffect in problems of the design of
  optimal digital control systems.
\newblock \emph{Automation and Remote Control}, 53\penalty0 (10):\penalty0
  1523--1528, 1992.

\bibitem[Louisell(2001)]{louisell2001delay}
James Louisell.
\newblock Delay differential systems with time-varying delay: New directions
  for stability theory.
\newblock \emph{Kybernetika}, 37\penalty0 (3):\penalty0 239--251, 2001.

\bibitem[Teel et~al.(1998)Teel, Nesic, and Kokotovic]{teel1998note}
Andrew~R Teel, Dragan Nesic, and Petar~V Kokotovic.
\newblock A note on input-to-state stability of sampled-data nonlinear systems.
\newblock In \emph{Decision and Control, 1998. Proceedings of the 37th IEEE
  Conference on}, volume~3, pages 2473--2478. IEEE, 1998.

\bibitem[Fridman and Shaked(2003)]{fridman2003delay}
Emilia Fridman and Uri Shaked.
\newblock Delay-dependent stability and $ h_\infty$ control: constant and
  time-varying delays.
\newblock \emph{International journal of control}, 76\penalty0 (1):\penalty0
  48--60, 2003.

\bibitem[Niculescu and Gu(2012)]{niculescu2012advances}
Silviu-Iulian Niculescu and Keqin Gu.
\newblock \emph{Advances in time-delay systems}, volume~38.
\newblock Springer Science \& Business Media, 2012.

\bibitem[Sun et~al.(1993)Sun, Nagpal, and Khargonekar]{sun1993h}
Weiqian Sun, Krishan~M Nagpal, and Pramod~P Khargonekar.
\newblock $ h_\infty$ control and filtering for sampled-data systems.
\newblock \emph{IEEE Transactions on Automatic Control}, 38\penalty0
  (8):\penalty0 1162--1175, 1993.

\bibitem[Kabamba and Hara(1993)]{kabamba1993worst}
Pierre~T Kabamba and Shinji Hara.
\newblock Worst-case analysis and design of sampled-data control systems.
\newblock \emph{IEEE Transactions on Automatic Control}, 38\penalty0
  (9):\penalty0 1337--1358, 1993.

\bibitem[Toivonen(1992)]{toivonen1992sampled}
Hannu~T Toivonen.
\newblock Sampled-data $ h_\infty$ optimal control of time-varying systems.
\newblock \emph{Automatica}, 28\penalty0 (4):\penalty0 823--826, 1992.

\bibitem[Dullerud and Lall(1999)]{dullerud1999asynchronous}
Geir~E Dullerud and Sanjay Lall.
\newblock Asynchronous hybrid systems with jumps-analysis and synthesis
  methods.
\newblock \emph{Systems \& Control Letters}, 37\penalty0 (2):\penalty0 61--69,
  1999.

\bibitem[Blondel and Nesterov(2005)]{blondel2005computationally}
Vincent~D Blondel and Yurii Nesterov.
\newblock Computationally efficient approximations of the joint spectral
  radius.
\newblock \emph{SIAM Journal on Matrix Analysis and Applications}, 27\penalty0
  (1):\penalty0 256--272, 2005.

\bibitem[Molchanov and Pyatnitskiy(1989)]{molchanov1989criteria}
Alexander~P Molchanov and Ye~S Pyatnitskiy.
\newblock Criteria of asymptotic stability of differential and difference
  inclusions encountered in control theory.
\newblock \emph{Systems \& Control Letters}, 13\penalty0 (1):\penalty0 59--64,
  1989.

\bibitem[Daafouz and Bernussou(2001)]{daafouz2001parameter}
Jamal Daafouz and Jacques Bernussou.
\newblock Parameter dependent lyapunov functions for discrete time systems with
  time varying parametric uncertainties.
\newblock \emph{Systems \& control letters}, 43\penalty0 (5):\penalty0
  355--359, 2001.

\bibitem[Lee(2006)]{lee2006uniform}
J-W Lee.
\newblock On uniform stabilization of discrete-time linear parameter-varying
  control systems.
\newblock \emph{IEEE Transactions on Automatic Control}, 51\penalty0
  (10):\penalty0 1714--1721, 2006.

\bibitem[Megretski(1996)]{megretski1996integral}
Alexandre Megretski.
\newblock Integral quadratic constraints derived from the set-theoretic
  analysis of difference inclusions.
\newblock In \emph{Decision and Control, 1996., Proceedings of the 35th IEEE
  Conference on}, volume~3, pages 2389--2394. IEEE, 1996.

\bibitem[Hu and Blanchini(2010)]{hu2010non}
Tingshu Hu and Franco Blanchini.
\newblock Non-conservative matrix inequality conditions for
  stability/stabilizability of linear differential inclusions.
\newblock \emph{Automatica}, 46\penalty0 (1):\penalty0 190--196, 2010.

\bibitem[van~de Wouw et~al.(2010)van~de Wouw, Naghshtabrizi, Cloosterman, and
  Hespanha]{van2010tracking}
Nathan van~de Wouw, Payam Naghshtabrizi, MBG Cloosterman, and Joao~Pedro
  Hespanha.
\newblock Tracking control for sampled-data systems with uncertain time-varying
  sampling intervals and delays.
\newblock \emph{International Journal of Robust and Nonlinear Control},
  20\penalty0 (4):\penalty0 387--411, 2010.

\bibitem[Cloosterman et~al.(2010)Cloosterman, Hetel, Van~de Wouw, Heemels,
  Daafouz, and Nijmeijer]{cloosterman2010controller}
Marieke~BG Cloosterman, Laurentiu Hetel, Nathan Van~de Wouw, WPMH Heemels,
  Jamal Daafouz, and Henk Nijmeijer.
\newblock Controller synthesis for networked control systems.
\newblock \emph{Automatica}, 46\penalty0 (10):\penalty0 1584--1594, 2010.

\bibitem[Kao and Lincoln(2004)]{kao2004simple}
Chung-Yao Kao and Bo~Lincoln.
\newblock Simple stability criteria for systems with time-varying delays.
\newblock \emph{Automatica}, 40\penalty0 (8):\penalty0 1429--1434, 2004.

\bibitem[Mirkin(2007)]{mirkin2007some}
Leonid Mirkin.
\newblock Some remarks on the use of time-varying delay to model
  sample-and-hold circuits.
\newblock \emph{IEEE Transactions on Automatic Control}, 52\penalty0
  (6):\penalty0 1109--1112, 2007.

\bibitem[Fridman et~al.(2004)Fridman, Seuret, and Richard]{fridman2004robust}
Emilia Fridman, Alexandre Seuret, and Jean-Pierre Richard.
\newblock Robust sampled-data stabilization of linear systems: an input delay
  approach.
\newblock \emph{Automatica}, 40\penalty0 (8):\penalty0 1441--1446, 2004.

\bibitem[Fujioka(2009)]{fujioka2009stability}
Hisaya Fujioka.
\newblock Stability analysis of systems with aperiodic sample-and-hold devices.
\newblock \emph{Automatica}, 45\penalty0 (3):\penalty0 771--775, 2009.

\bibitem[Omran et~al.(2014)Omran, Hetel, Richard, and
  Lamnabhi-Lagarrigue]{omran2014stability}
Hassan Omran, Laurentiu Hetel, Jean-Pierre Richard, and Fran{\c{c}}oise
  Lamnabhi-Lagarrigue.
\newblock Stability analysis of bilinear systems under aperiodic sampled-data
  control.
\newblock \emph{Automatica}, 50\penalty0 (4):\penalty0 1288--1295, 2014.

\bibitem[Velasco et~al.(2003)Velasco, Fuertes, and Marti]{velasco2003self}
Manel Velasco, Josep Fuertes, and Pau Marti.
\newblock The self triggered task model for real-time control systems.
\newblock In \emph{Work-in-Progress Session of the 24th IEEE Real-Time Systems
  Symposium (RTSS03)}, volume 384, 2003.

\bibitem[Wang and Lemmon(2008)]{wang2008state}
Xiaofeng Wang and Michael Lemmon.
\newblock State based self-triggered feedback control systems with l2
  stability.
\newblock In \emph{17th IFAC world congress}, 2008.

\bibitem[Anta and Tabuada(2008)]{anta2008self}
Adolfo Anta and Paulo Tabuada.
\newblock Self-triggered stabilization of homogeneous control systems.
\newblock In \emph{American Control Conference, 2008}, pages 4129--4134. IEEE,
  2008.

\bibitem[Anta and Tabuada(2010)]{anta2010sample}
Adolfo Anta and Paulo Tabuada.
\newblock To sample or not to sample: Self-triggered control for nonlinear
  systems.
\newblock \emph{IEEE Transactions on Automatic Control}, 55\penalty0
  (9):\penalty0 2030--2042, 2010.

\bibitem[Tabuada(2007)]{tabuada2007event}
Paulo Tabuada.
\newblock Event-triggered real-time scheduling of stabilizing control tasks.
\newblock \emph{IEEE Transactions on Automatic Control}, 52\penalty0
  (9):\penalty0 1680--1685, 2007.

\bibitem[Mazo et~al.(2009)Mazo, Anta, and Tabuada]{mazo2009self}
Manuel Mazo, Adolfo Anta, and Paulo Tabuada.
\newblock On self-triggered control for linear systems: Guarantees and
  complexity.
\newblock In \emph{Control Conference (ECC), 2009 European}, pages 3767--3772.
  IEEE, 2009.

\bibitem[Mazo et~al.(2010)Mazo, Anta, and Tabuada]{mazo2010iss}
Manuel Mazo, Adolfo Anta, and Paulo Tabuada.
\newblock An iss self-triggered implementation of linear controllers.
\newblock \emph{Automatica}, 46\penalty0 (8):\penalty0 1310--1314, 2010.

\bibitem[Wang and Lemmon(2009)]{wang2009self}
Xiaofeng Wang and Michael~D Lemmon.
\newblock Self-triggered feedback control systems with finite-gain $ l_2 $
  stability.
\newblock \emph{IEEE transactions on automatic control}, 54\penalty0
  (3):\penalty0 452--467, 2009.

\bibitem[Wang and Lemmon(2010)]{wang2010self}
Xiaofeng Wang and Michael~D Lemmon.
\newblock Self-triggering under state-independent disturbances.
\newblock \emph{IEEE Transactions on Automatic Control}, 55\penalty0
  (6):\penalty0 1494--1500, 2010.

\bibitem[Lemmon et~al.(2007)Lemmon, Chantem, Hu, and Zyskowski]{lemmon2007self}
Michael Lemmon, Thidapat Chantem, Xiaobo Hu, and Matthew Zyskowski.
\newblock On self-triggered full-information h-infinity controllers.
\newblock \emph{Hybrid Systems: computation and control}, pages 371--384, 2007.

\bibitem[Khalil(2002)]{khalil2002nonlinear}
Hassan~K Khalil.
\newblock Nonlinear systems, 3rd.
\newblock \emph{New Jewsey, Prentice Hall}, 9\penalty0 (4.2), 2002.

\bibitem[Fiter et~al.(2012)Fiter, Hetel, Perruquetti, and
  Richard]{fiter2012state}
Christophe Fiter, Laurentiu Hetel, Wilfrid Perruquetti, and Jean-Pierre
  Richard.
\newblock A state dependent sampling for linear state feedback.
\newblock \emph{Automatica}, 48\penalty0 (8):\penalty0 1860--1867, 2012.

\end{thebibliography}

\section{Appendix: Proofs}\label{Appendix}
\begin{proof}[Proof of Proposition \ref{STMS-off-p}:]
	Consider $ x_k\in\mathbb{R}^n $	and the quadratic Lyapunov function $  V(x)=x^T P x $ in which the matrix $ P $ satisfies \eqref{LMI1-STM-on-p} and \eqref{LMI2-STM-on-p}.	
	\begin{enumerate}
		\item [Case (i):]  $ x_k \not\in \mathcal{E}(P,1)  $. Firstly we should prove that the set $ \bar{S}_{l_{min}}^{l_{max}}(\Gamma,x_k) $ in not empty (i.e. for a given $ x_k  \in \mathbb{R}^n $, $\exists \sigma \in {\bar{S}}^{l_{max}}_{l_{min}}(\Gamma,x_k) $). From \eqref{LMI2-STM-on-p} and Schur complement, we have $ M=M^T\succeq 0 $, and $ \frac{\gamma}{\chi}I-PM^{-1}P- P \succeq 0 $. The latter equivalently can be written as 
		\begin{equation}\label{equival-LMI2}
			\frac{\gamma}{\chi} \geq \lambda_{max} \big(PM^{-1}P+P\big).   
		\end{equation}
		consider a sampling horizon $ \sigma^* $ satisfying \eqref{LMI1-STM-on-p}. Then, from \eqref{equival-LMI2} and \eqref{U_sigma}, one has 	
		\begin{equation}
			\begin{pmatrix}
				x_k \\
				1
			\end{pmatrix}^T U_{\sigma^*} \begin{pmatrix}
				x_k \\
				1
			\end{pmatrix} \geq 0,
		\end{equation}	
		which implies that $ \sigma^* \in {\bar{S}}^{l_{max}}_{l_{min}}(\Gamma,x_k) $ (i.e. the set $\bar{S}_{l_{min}}^{l_{max}}(\Gamma,x_k) $ in not empty, for any $ x_k \notin \mathcal{E}(P,1)$).
		
		Now, consider $ \sigma_k \in \bar{S}_{l_{min}}^{l_{max}}(\Gamma,x_k) $. Then, we have	
		\begin{equation}\label{NinRP2-on}
			\begin{split}
				V(x_{k+1}) & = x_{k+1}^TPx_{k+1} \\
				& = \big(\Phi_{\sigma_k} x_k+\bar{w}_k\big)^TP\big(\Phi_{\sigma_k} x_k+\bar{w}_k\big) \\
				&= x^T_k \Phi^T_{{\sigma}_k} P \Phi_{{\sigma}_k} x_k+ x^T_k \Phi^T_{{\sigma}_k} P \bar{w}_k \\
				& \quad +\bar{w}^T_k P \Phi_{{\sigma}_k} x_k+\bar{w}^T_k P \bar{w}_k .
			\end{split}
		\end{equation}
		
		Using the inequality $ x^Ty+y^Tx \leq x^T M x+ y^T M^{-1} y $ and for any $ x\in \mathbb{R}^n $, $ y\in \mathbb{R}^n $, and $ M=M^T\succ 0 $, we have	
		\begin{equation}\label{inq}
			\begin{split}
				x^T_k \Phi^T_{{\sigma}_k} P \bar{w}_k+\bar{w}^T_k P \Phi_{{\sigma}_k} x_k & \leq x^T_k \Phi^T_{{\sigma}_k} M \Phi_{{\sigma}_k} x_k \\  
				& \quad + \bar{w}^T_k P M^{-1} P \bar{w}_k. 
			\end{split}
		\end{equation} 
		
		Therefore, from \eqref{NinRP2-on}, we have 
		\begin{equation}\label{NinRP2-onn}
			\begin{split}
				V(x_{k+1}) &= x^T_k \Phi^T_{{\sigma}_k} P \Phi_{{\sigma}_k} x_k+ x^T_k \Phi^T_{{\sigma}_k} P \bar{w}_k \\
				& \quad +\bar{w}^T_k P \Phi_{{\sigma}_k} x_k +\bar{w}^T_k P \bar{w}_k \\
				& \leq x^T_k \Phi^T_{{\sigma}_k}\Big( P + M \Big) \Phi_{{\sigma}_k} x_k \\
				& \quad   + \bar{w}^T_k \Big( P M^{-1} P + P \Big) \bar{w}_k ,
			\end{split}
		\end{equation}
		and furthermore
		\begin{equation}\label{NinRPLMI-on}
			\begin{split}
				V(x_{k+1}) \leq
				\begin{pmatrix}
					x_k\\
					1
				\end{pmatrix}^T \Lambda_{\sigma_k} \begin{pmatrix}
					x_k\\
					1
				\end{pmatrix},
			\end{split}
		\end{equation}
		in which
		\begin{equation}\label{Lambda-0n}
			\begin{split}
				\Lambda_{\sigma_k}  = \begin{pmatrix}
					{\Lambda_{\sigma_k}}_{11} 
					& \textbf{0}\\
					\textbf{0} & {\Lambda_{\sigma_k}}_{22} 
				\end{pmatrix}.
			\end{split}
		\end{equation}
		where ${\Lambda_{\sigma_k}}_{11} = \Phi^T_{{\sigma}_k} \big(P+M\big)\Phi_{{\sigma}_k}$ and ${\Lambda_{\sigma_k}}_{22} = \Big(\varpi  \sum_{q=0}^{|\sigma_k|-1}C^q \Big)^2\lambda_{max}(P M^{-1} P+P) $.
		
		Then, since $ x_k \notin \mathcal{E}(P,1) $ (i.e. $ x_k^T P x_k > 1 $), we have	
		\begin{equation}\label{NinRPLMION}
			\begin{split}
				V(x_{k+1}) & \leq
				\begin{pmatrix}
					x_k\\
					1
				\end{pmatrix}^T \Bigg[\Lambda_{\sigma_k}+ \begin{bmatrix}
					-\gamma P & \textbf{0}\\
					\textbf{0} & \gamma
				\end{bmatrix} \Bigg] \begin{pmatrix}
					x_k\\
					1
				\end{pmatrix},
			\end{split}
		\end{equation}	
		with $ \gamma>0 $. Then, from \eqref{es-on-p}, we get		
		\begin{equation}
			\begin{split}	
				V(x_{k+1}) & \leq \begin{pmatrix}
					x_k\\
					1
				\end{pmatrix}^T 
				\begin{pmatrix}
					e^{(-\beta\sum\nolimits_{j=1}^{|{\sigma_k} |}T_{\sigma_k}^{j }) }P-\gamma P & \textbf{0}\\
					\textbf{0} & \gamma
				\end{pmatrix}
				\begin{pmatrix}
					x_k\\
					1
				\end{pmatrix}\\
				& \leq e^{(-\beta\sum\nolimits_{j=1}^{|{\sigma_k} |}T_{\sigma_k}^{j }) } x_k^T P x_k + \gamma (1-x_k^T P x_k),
			\end{split}
		\end{equation}
		Therefore, since $x_k^T P x_k >  1 $, one has $ V(x_{k+1}) \leq e^{(-\beta\sum\nolimits_{j=1}^{|{\sigma_k} |}T_{\sigma_k}^{j }) } V(x_k)$. 	
		\item [ Case (ii):]  $ x_k \in \mathcal{E}(P,1) $. According to the self-triggering mechanism \eqref{STM-on-p} and from \eqref{CLSP}, we have	
		\begin{equation}
			x_{k+1} = \Phi_{\sigma_k}x_k+\bar{w}_k,
		\end{equation}
		where $ \sigma_k = \big(T_{max}\big) $ in which $ T_{max}=\max \{T\in\Gamma\} $. Then, $ \Phi_{\sigma_k}=\tilde{A}_{(T_{max})} $ and $ \tau_{k+1}=\tau_k+T_{max} $. From the section \ref{sysref}, one has
		\begin{equation}\label{interball1}
			\begin{split}
				\Big\| x_{k+1} \Big\|_2 \leq \Big\| \Phi_{\sigma_k}x_k \Big\|_2+\Big\| \bar{w}_k \Big\|_2 & \leq C'\Big\|  x_k \Big\|_2 + \varpi \\
				& \leq \frac{C'}{\lambda_{min}(P)}+ \varpi=\sqrt{\eta}.
			\end{split}
		\end{equation}
		where $ C'=\Big\| \Phi_{\sigma_k}\Big\|_2=\Big\|\tilde{A}_{(T_{max})} \Big\|_2 $  and $ \varpi $ is given in Assumption \ref{pertbounded}. We want to find a scalar $ \mu >0 $ such that  $  \mathcal{B}(0,\eta) \subset \mathcal{E}(P,\mu)  $. We know that $ x_k^T P x_k \leq \lambda_{max}(P) x_k^T x_k $, then form \eqref{interball1},
		
		\begin{equation}
			\mu \leq  \lambda_{max}(P) \Big(\frac{C'}{\lambda_{min}(P)}+ \varpi\Big)^2.
		\end{equation} 
		
	\end{enumerate}
	From  (i) and  (ii), one has $ x_k \in \mathcal{E}(P,\mu),~k\in\mathbb{N} $. Therefore, the solution of the system \eqref{CLSP} is globally uniformly ultimately bounded with the proposed self triggering mechanisms \eqref{STM-on-p}.
\end{proof}
\begin{proof}[Proof of Proposition \ref{STM-off-p}:]
	
	Consider $ x_k\in\mathbb{R}^n $	and the quadratic Lyapunov function $  V(x)=x^T P x $ in which the matrix $ P $ satisfies \eqref{SigmaOmega-off}.	
	\begin{enumerate}
		\item [Case (i):]  $ x_k \not\in \mathcal{E}(P,1)  $. Firstly we should prove that the set $ \bar{S}_{l_{min}}^{l_{max}}(\Gamma,x_k) $ in not empty (i.e. there exists at least the horizon $ \sigma^* \in {\bar{S}}^{l_{max}}_{l_{min}}(\Gamma,x_k),~ \forall x_k \in \mathbb{R}^n $). From \eqref{es-off-p}, the matrix $ U_c $ can be written as
		
		\begin{equation}
			U_c=U+\begin{pmatrix}
				\epsilon_c Q_c & * & *\\
				\textbf{0} & \textbf{0} & *\\
				\textbf{0} & \textbf{0} &  0 
			\end{pmatrix},
		\end{equation}
		which is composed of two semi-definite matrices. Then, clearly, it can be concluded that $ U_c $ is also a semi-definite matrix. Therefore, there exists at least the horizon $ \sigma^* \in {\bar{S}}^{l_{max}}_{l_{min}}(\Gamma,x_k),~ \forall x_k \in \mathbb{R}^n $. 
		
		Now, consider $ \sigma_k \in \bar{S}_{l_{min}}^{l_{max}}(\Gamma,x_k) $,		
		We want to guarantee that 
		\begin{equation}\label{NinRP}
			\begin{split}
				x_{k+1}^TPx_{k+1}<\bar{\beta} x_{k}^TPx_{k},~\bar{\beta}=e^{(-\beta\sum\nolimits_{j=1}^{|{\sigma} |}T_{\sigma}^{j }) }
			\end{split}
		\end{equation}
		
		From \eqref{CLSP} and \eqref{NinRP}, we have
		\begin{equation}\label{NinRP2}
			\begin{split}
				&\big(\Phi_{\sigma_k} x_k+\bar{w}_k\big)^TP\big(\Phi_{\sigma_k} x_k+\bar{w}_k\big) \leq \bar{\beta} x_{k}^TPx_{k}\\
				& x^T_k \Phi^T_{{\sigma}_k} P \Phi_{{\sigma}_k} x_k+ x^T_k \Phi^T_{{\sigma}_k} P \bar{w}_k \\
				& \qquad +\bar{w}^T_k P \Phi_{{\sigma}_k} x_k+\bar{w}^T_k P \bar{w}_k \leq-\bar{\beta} x_{k}^TPx_{k},
			\end{split}
		\end{equation}
		equivalently, equation \eqref{NinRP2} can be written as 
		\begin{equation}\label{NinRPLMI}
			\begin{split}
				\begin{pmatrix}
					x_k\\
					\bar{w}_k\\
					1
				\end{pmatrix}^T \Lambda_{\sigma_k} \begin{pmatrix}
					x_k\\
					\bar{w}_k\\
					1
				\end{pmatrix}>0,
			\end{split}
		\end{equation}
		in which
		\begin{equation}\label{Lambda}
			\begin{split}
				\Lambda_{\sigma_k} = \begin{pmatrix}
					-\Phi_{\sigma_k}^TP\Phi_{\sigma_k}+\bar{\beta}P & * & *\\
					-P\Phi_{\sigma_k} & -P & *\\
					\textbf{0} & \textbf{0} & 0
				\end{pmatrix}.
			\end{split}
		\end{equation}
		
		Since $ \Big\|\bar{w}_k\Big\|_2\leq  \varpi \sum_{q=0}^{|\sigma_k|-1}C^q=\theta $, this can be shown as 
		\begin{equation}\label{PertLMI}
			\begin{split}
				\begin{pmatrix}
					x_k\\
					\bar{w}_k\\
					1
				\end{pmatrix}^T      
				\begin{pmatrix}
					\textbf{0} & \textbf{0} & \textbf{0}\\
					\textbf{0} & \frac{-1}{ \theta}I & \textbf{0}\\
					\textbf{0} & \textbf{0} & 1
				\end{pmatrix}
				\begin{pmatrix}
					x_k\\
					\bar{w}_k\\
					1
				\end{pmatrix}\geq 0.
			\end{split}
		\end{equation}
		
		Also, the constraint $ x(k)\not\in \mathcal{E}(P,1) $ can be written as
		\begin{equation}\label{NILMI}
			\begin{split}
				\begin{pmatrix}
					x_k\\
					\bar{w}_k\\
					1
				\end{pmatrix}^T      
				\begin{pmatrix}
					P & \textbf{0} & \textbf{0}\\
					\textbf{0} & \textbf{0} & \textbf{0}\\
					\textbf{0} & \textbf{0} & -1
				\end{pmatrix}
				\begin{pmatrix}
					x_k\\
					\bar{w}_k\\
					1
				\end{pmatrix}\geq 0.
			\end{split}
		\end{equation}
		
		From equations \eqref{NinRPLMI},\eqref{PertLMI} and \eqref{NILMI} and by using S-procedure, the aim is to guarantee that
		\begin{equation}\label{FLMI}
			\begin{split}
				&\begin{pmatrix}
					x_k\\
					\bar{w}_k\\
					1
				\end{pmatrix}^T \times \Lambda'_{\sigma_k}  \times \begin{pmatrix}
					x_k\\
					\bar{w}_k\\
					1
				\end{pmatrix}\geq 0.
			\end{split}
		\end{equation}
		in which $ \gamma_1, \gamma_2 > 0 $ and \\ $\Lambda'_{\sigma_k} = \begin{pmatrix}
			-\Phi_{\sigma_k}^TP\Phi_{\sigma_k}+(\bar{\beta}-\gamma_1) P & * & *\\
			-P\Phi_{\sigma_k} & \frac{\gamma_2}{ \theta}I-P & *\\
			\textbf{0} & \textbf{0} & {-\gamma_2+\gamma_1}
		\end{pmatrix}$.
		
		Let $ x_{k} \in \mathbb{R}^n, k\in \mathbb{N} $, There exists a conic region $ \mathcal{R}_c, c \in \mathbb{N} $ as in \eqref{conicreg} such that $ x_{k} \in  \mathcal{R}_c$. Using the S-procedure, one can obtain that the inequality \eqref{FLMI} is satisfied if and only if there exists a scalar $ \epsilon_c >0 $ such that
		\begin{equation}\label{LMI-FLMI}
			\begin{split}
				\begin{pmatrix}
					\epsilon_c Q_c-\Phi_{\sigma_k}^TP\Phi_{\sigma_k}+(\bar{\beta}-\gamma_1) P & * & *\\
					-P\Phi_{\sigma_k} & \frac{\gamma_2}{\theta}I-P & *\\
					\textbf{0} & \textbf{0} &  {-\gamma_2+\gamma_1}
				\end{pmatrix} \\ 
				\succeq 0
			\end{split}
		\end{equation}
		
		Therefore, one has $ V(x_{k+1}) \leq e^{(-\beta\sum\nolimits_{j=1}^{|{\sigma_k} |}T_{\sigma_k}^{j }) } V(x_k)$ for every $ x_k \not\in \mathcal{E}(P,1)  $. 
		\item [Case (ii):]  $ x_k \in \mathcal{E}(P,1) $. According to the self-triggering mechanism \eqref{STM-on-p} and from \eqref{CLSP}, we have	
		\begin{equation}
			x_{k+1} = \Phi_{\sigma_k}x_k+\bar{w}_k,
		\end{equation}
		where $ \sigma_k = \big(T_{max}\big) $ in which $ T_{max}=\max \{T\in\Gamma\} $. Then, $ \Phi_{\sigma_k}=\tilde{A}_{(T_{max})} $ and $ \tau_{k+1}=\tau_k+T_{max} $. From the section \ref{sysref}, one has
		\begin{equation}\label{interball}
			\begin{split}
				\Big\| x_{k+1} \Big\|_2 \leq \Big\| \Phi_{\sigma_k}x_k \Big\|_2+\Big\| \bar{w}_k \Big\|_2 & \leq C'\Big\|  x_k \Big\|_2 + \varpi \\
				& \leq \frac{C'}{\lambda_{min}(P)}+ \varpi=\sqrt{\eta}.
			\end{split}
		\end{equation}
		where $ C'=\Big\| \Phi_{\sigma_k}\Big\|_2=\Big\|\tilde{A}_{(T_{max})} \Big\|_2 $  and $ \varpi $ is given in Assumption \ref{pertbounded}. We want to find a scalar $ \mu >0 $ such that  $  \mathcal{B}(0,\eta) \subset \mathcal{E}(P,\mu)  $. We know that $ x_k^T P x_k \leq \lambda_{max}(P) x_k^T x_k $, then form \eqref{interball},
		
		\begin{equation}
			\mu \leq  \lambda_{max}(P) \Big(\frac{C'}{\lambda_{min}(P)}+ \varpi\Big)^2.
		\end{equation}  
	\end{enumerate}
	From  (i) and  (ii), one has $ x_k \in \mathcal{E}(P,\mu),~k\in\mathbb{N} $. Therefore, the solution of the system \eqref{CLSP} is globally uniformly ultimately bounded with the proposed self triggering mechanisms \eqref{STMoff-p}.  
\end{proof}

\end{document}